\documentclass[fleqn,usenatbib]{mnras}

\usepackage{newtxtext,newtxmath}

\usepackage[T1]{fontenc}

\DeclareRobustCommand{\VAN}[3]{#2}
\let\VANthebibliography\thebibliography
\def\thebibliography{\DeclareRobustCommand{\VAN}[3]{##3}\VANthebibliography}

\usepackage{graphicx}	
\usepackage{amsmath}	

\title[Saltire]{Saltire - A model to measure dynamical masses for high-contrast binaries and exoplanets with high-resolution spectroscopy.}

\author[D. Sebastian et al.]{
D. Sebastian$^{1}$ $^{\href{https://orcid.org/0000-0002-2214-9258}{\includegraphics[scale=0.5]{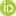}}}$\thanks{E-mail: D.Sebastian.1@bham.ac.uk}, 
A. H.M.J. Triaud$^{1}$ $^{\href{https://orcid.org/0000-0002-5510-8751}{\includegraphics[scale=0.5]{Images/orcid.jpg}}}$,
M. Brogi$^{2,3,4}$ $^{\href{https://orcid.org/0000-0002-7704-0153}{\includegraphics[scale=0.5]{Images/orcid.jpg}}}$\\
$^{1}$ School of Physics \& Astronomy, University of Birmingham, Edgbaston, Birmimgham, B15 2TT, UK \\
$^{2}$ Dipartimento di Fisica, Università degli Studi di Torino, via Pietro Giuria 1, I-10125 Torino, Italy\\
$^{3}$ Department of Physics, University of Warwick, Coventry CV4 7AL, UK\\
$^{4}$ INAF-Osservatorio Astrofisico di Torino, Via Osservatorio 20, I-10025 Pino Torinese, Italy\\
}
\date{Accepted. Received; in original form}

\pubyear{2023}

\begin{document}
\label{firstpage}
\pagerange{\pageref{firstpage}--\pageref{lastpage}}
\maketitle

\begin{abstract} 
High-resolution cross-correlation methods are widely used to discover and to characterise atomic and molecular species in exoplanet atmospheres. The characteristic cross-correlation signal is typically represented as a function of the velocity of the system, and the semi-amplitude of the planet's orbit.
We present {\tt Saltire}, a fast and simple model that accurately reproduces the shape of such cross-correlation signals, allowing a direct fit to the data by using a minimum set of parameters. We show how to use this model on the detection of atmospheric carbon monoxide in archival data of the hot Jupiter $\rm \tau$ Bo\"otis\,b, and how {\tt Saltire} can be used to estimate the semi-amplitude and rest velocity of high brightness-ratio binaries.
By including the shape of the signal, we demonstrate that our model allows to robustly derive the signal position up to 10 times more accurate, compared to conventional methods. Furthermore, we discuss the impact of correlated noise and demonstrate that {\tt Saltire} is a robust tool for estimating systematic uncertainties on the signal position.
{\tt Saltire} opens a new door to analyse high signal-to-noise data to accurately study atmospheric dynamics and to measure precise dynamical masses for exoplanets and faint stellar companions. We show, that the phase-resolved shape of the atmospheric cross-correlation signal can accurately be reproduced, allowing studies of phase-dependent signal changes and to disentangle them from noise and data aliases.

\end{abstract}




\section{Introduction}

In the past decades, many tools have been developed to characterise planetary atmospheres using ground and space-based spectro-photometry, as well as ground-based high resolution spectroscopy. One important and widely used method to detect exoplanet atmospheres is High-Resolution Cross-Correlation Spectroscopy \citep[HRCCS;][]{snellen2010} which has been used to make robust detections of atomic and molecular species within exoplanet atmospheres, and used to produce empirical information about exoplanet atmospheric dynamics. An advantage of this method is that it is not limited to transiting exoplanets.

Using the HRCCS method, planetary atmospheric signals are retrieved by obtaining consecutive spectroscopic measurements of high resolution, either in transmission during the planet's transit or close to its superior conjunction to measure the planet's day-side thermal emission \citep[see][for a review]{Birkby2018}. The significance of a detection mainly depends on two aspects: (1) the signal's strength grows with the square-root of the number of atmospheric lines that have been correlated, thus favouring spectral observations with wider wavelength range and/or higher spectral resolutions; (2) the signal's amplitude is a function of the contrast ratio of the exoplanet's atmosphere to the stellar photosphere, thus favouring close-in gas giants with high equilibrium temperatures ($T_{\rm eq} \geq 2200\,\rm K$), so called ultra-hot Jupiters \citep[UHJs;][]{Parmentier2018}. A crucial aspect of the HRCCS method is that spectra are corrected from telluric and stellar contributions which is generally possible because the planet signal is moving fast relative to them (by several $\rm km\,s^{-1}$) within the span of a typical observation. The planetary signal is thus not correlated to the spectrograph or the stellar rest frame. There are several different methods for telluric and stellar line fitting \citep[e.g.][]{Brogi12,Brogi16}, or singular value decomposition \citep{kalman1996} that have been applied successfully to many exoplanets and instrumental setups \citep[e.g.][]{deKok13,Birkby13,Piskorz16,Cheverall23}.

\subsection{$ K_{\rm c} - V_{\rm rest}$ mapping} 
In the HRCCS framework, spectra corrected for telluric and stellar contributions are cross-correlated with a model spectrum of the planet's atmosphere or with a model of a specific atomic or molecular species, moved to the planet's rest frame. 
This method requires the semi-amplitude ($K_{\rm c}$) of the planet's orbit to be known exactly, which is usually not the case. Typically a cross-correlation map in the $K_{\rm c} - V_{\rm rest}$ plane is generated (`CCF map' hereafter), which probes the cross-correlation signal for several possible semi-amplitudes and rest velocities ($V_{\rm rest}$) \citep[e.g.][]{Brogi12,deKok13}.

Detecting the planet's atmospheric signal in the $K_{\rm c} - V_{\rm rest}$ plane using HRCCS thus literately turns the planet and its host star into a double-lined spectroscopic binary as the planet's semi-amplitude can be measured. This allows to measure the mass-ratio between host star and planet. For transiting planets this allows even to model independent, dynamical planet masses \citep{Birkby2018}.

Unfortunately, in these maps, $K_{\rm c}$ is often highly correlated to $V_{\rm rest}$ which can cause large uncertainties of the planet's semi-amplitude. Typical uncertainties of $K_{\rm c}$ are in the order of tens of kilometres \citep[e.g.][]{Brogi18, BelloArufe22, Yan2023}.
Furthermore it has been suggested that different species in the planet's atmosphere show different semi-amplitudes and rest velocities, different to the systemic velocity, both in infrared \citep{Brogi2023} and optical regimes \citep[e.g.][]{Nugroho20}. Atmospheric models reveal that the position and shape of the atmospheric signal in the CCF map can be used to disentangle the temperature structure, rotation, and dynamics of a planet's atmosphere, especially during transit observations \citep[e.g.][]{Wardenier2021,Kesseli22}.
Thus, characterising the accurate position of the planetary signal within the CCF map is essential to empirically understand these processes and essential to place constraints on atmospheric models for close-in giant planets.

Typically the planet's semi-amplitude and rest velocity are determined either by fitting a Gaussian to the cross-correlation map at the maximum significance level \citep[e.g.][]{Prinoth22}, or by statistically analysing the maximum likelihood of the detection \citep[e.g.][]{Brogi2019,Brogi2023}. 
Detections with state-of-the-art instrumentation have a generally low detection significance (typically $<10\,{\rm \sigma}$). Future instruments on large telescopes - like the ELT - will result in a higher detection significance of atmospheric signals in the CCF maps and thus will allow to disentangle the planet's orbit from dynamics in the atmosphere.

\subsection{Binaries as targets of opportunity}

A typical hot Jupiter in a very close orbit would have a contrast ratio in the order of $10^{-4}$ to $10^{-5}$ between the planetary atmosphere and the stellar photosphere. In \cite{Sebastian23a} we showed that eclipsing binaries of F,G, \& K-type stars, orbited by late type M-dwarf companions can be used to test the achievable accuracy of the the HRCCS method that is required to derive dynamical masses. First, the contrast ratio between both stars is $10^{-3}$ to $10^{-4}$, which is comparable to exoplanet atmospheres, making such binaries good proxies to exoplanets. Second, the spectrum of an M-star is well known, and we can use optimised model-spectra, observed high-resolution spectra, or even line-masks - optimised for M-dwarfs - as templates to generate CCF maps.  

A large sample of hundreds of Eclipsing Binaries with Low Mass (EBLM) companions, originally detected by the WASP survey \citep{pollacco06} have been spectroscopically observed in the EBLM project \citep{triaud13,triaud17}. The faint secondary implies the binary is a single-lined binary. This allows to use stabilised spectrographs and line lists, just like for single stars, to derive very precise radial velocities of the primary. The goal of this project is to derive precise orbital parameters from radial velocity (RV) observations in order to measure fundamental parameters for the secondary stars such as precise mass, and radius of low-mass M-dwarfs \citep[e.g.][]{boetticher19,Gill19,swayne21,Sebastian23}. As single-lined systems, only parameters relative to the primary can be obtained, but as double-lined systems, which the HRCCS method gives access to, absolute parameters can be measured. In addition these binaries are monitored to search for circumbinary exoplanets \citep{Triaud22, Standing23}. We note that binaries of any type, with large brightness differences can benefit from our approach (e.g. Sun-like companions to red giants). 

In this paper, we explore how to model the shape of a CCF map from a typical HRCCS observation. We introduce the {\tt Saltire} model, a versatile model that allows (i) to fit a HRCCS CCF map and determine accurate parameters for $K_{\rm c}$ and $V_{\rm rest}$ and (ii) to predict CCF maps from a HRCCS observation of exoplanet atmospheres.

\begin{figure}
    \includegraphics[width=\linewidth]{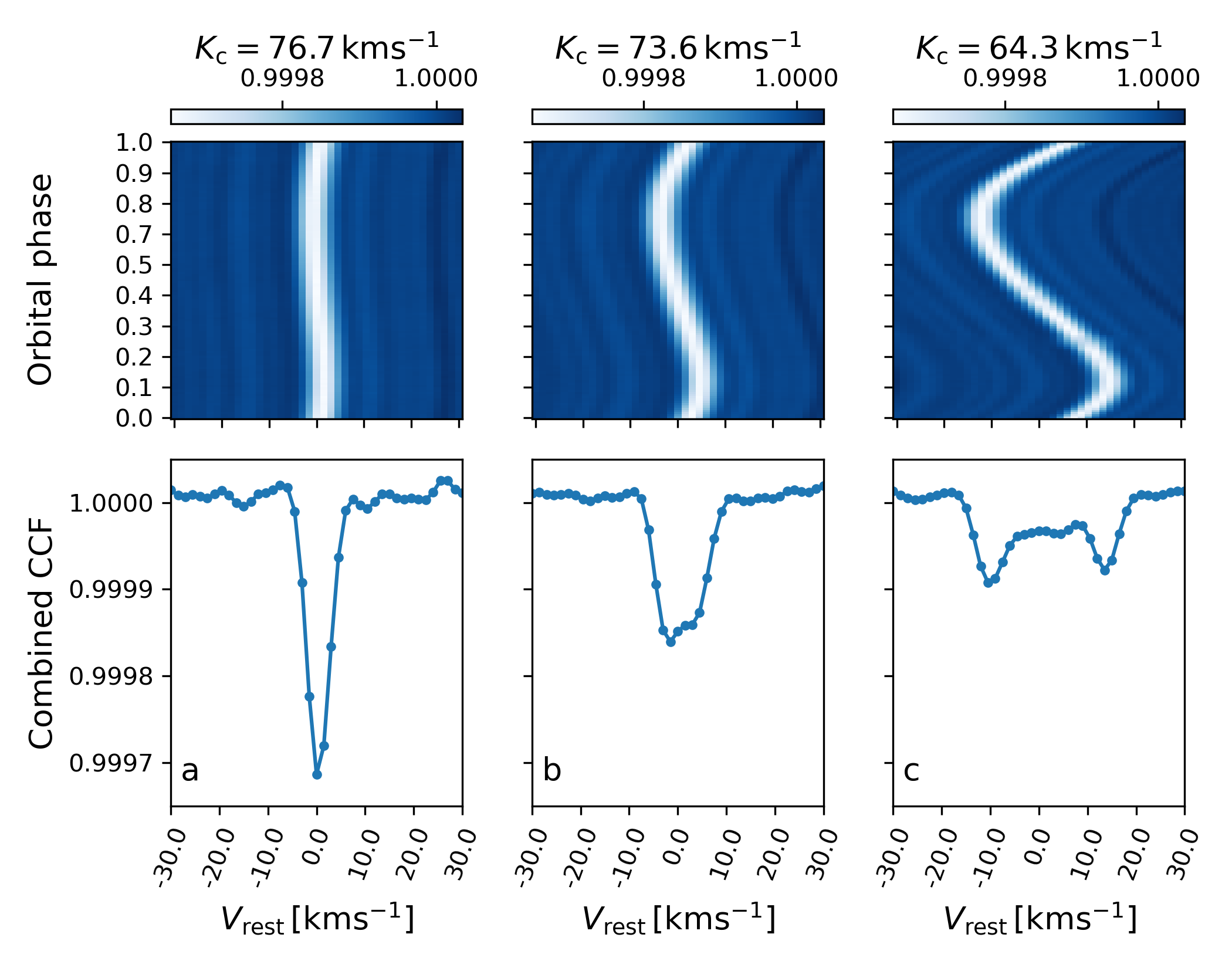}
    \caption{Upper panel: CCF functions for spectra, shifted to match different semi-amplitudes of the secondary. Lower panel: Combined CCFs for each semi-amplitude.}
    \label{fig:CCF_creation}
\end{figure}

\section{Modelling HRCCS observations} \label{model}

To model a CCF map, we first need to understand how it is created. It is a superposition of all spectroscopic observations which (i) have been post-processed by correcting for telluric and stellar lines from the host star and (ii) have been cross-correlated by a template spectrum. This is similarly true when analysing faint companions for EBLM binaries as it is for exo-atmospheric observations in transmission or day-side emission spectroscopy. The shape of this superposition depends on two main effects: (i) Changes of the relative contribution of the atmosphere for each measurement. This can be caused by intrinsic processes e.g. due to secondary eclipses, phase curves, or temperature gradients during transit observations, but also due to changing observing conditions or artefacts related to post processing of the data, like phase-depended stretching of the atmospheric signal due to the telluric and host star removal processes. (ii) The orbital reflex motion of the companion, which is moving on a Keplerian orbit.  

In a first step we can assume negligible changes in the relative contribution. This is particularly true for high-contrast  binaries where the companion is a M-dwarf, and where the phase dependent contribution of its absorption spectrum can be assumed to be constant and known to a first order (neglecting stellar variability, reflected light and ellipsoidal contributions for now). In this case the superposition of all cross-correlated spectra will reach a maximum CCF signal in the companion's rest frame, hence when all spectra have been corrected according to the companion's reflex motion. Typically in the HRCCS method the orbit parameters are assumed to be fixed (neglecting uncertainties of the orbital parameter determination). In case of close-in giant planets this is even simplified to a fixed sinusoidal motion.


\begin{figure}
    \includegraphics[width=\linewidth]{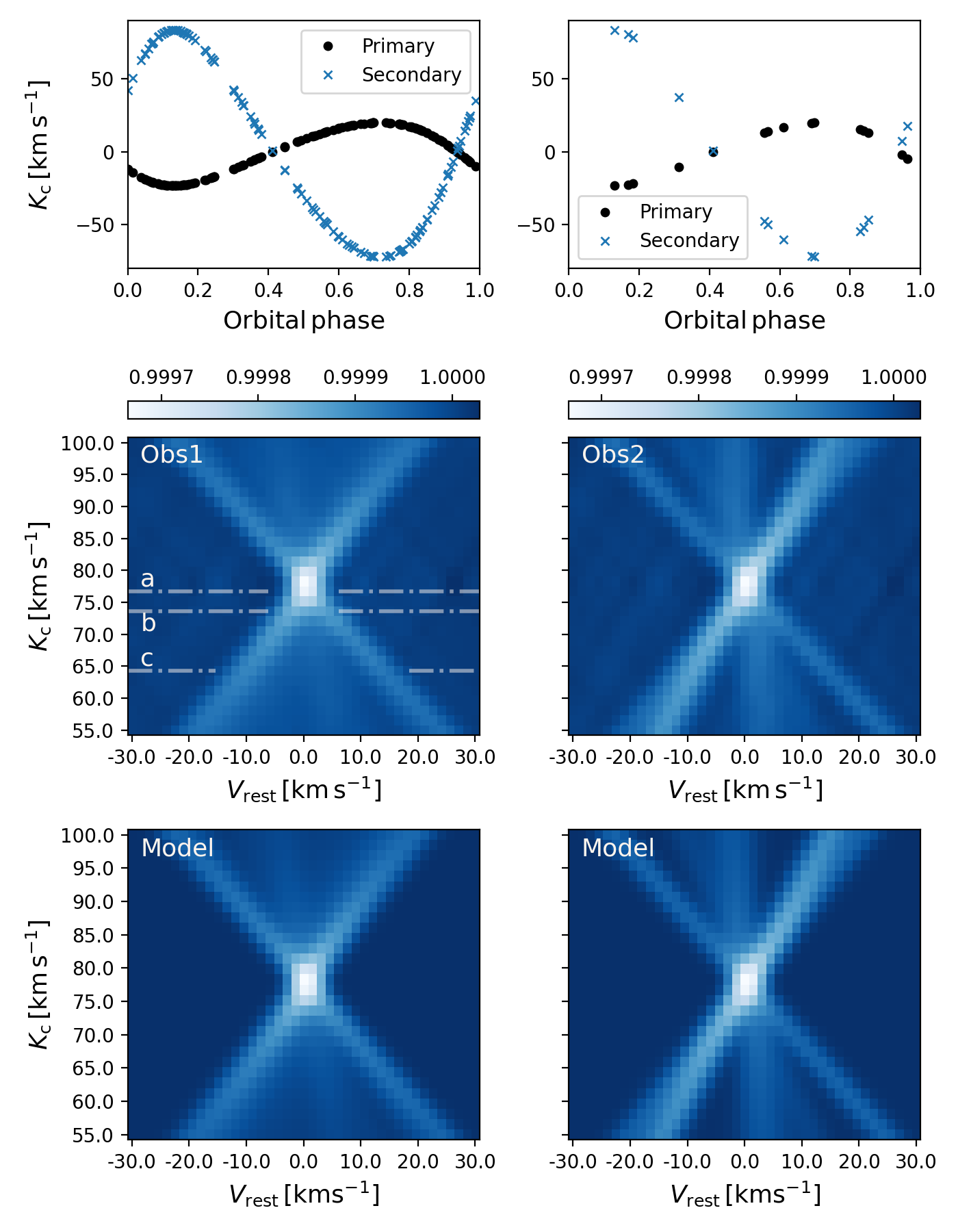}
    \caption{CCF map shapes from simulated EBLM observations and their dependence from phase coverage. Upper panel: simulated reflex motion of an EBLM binary. Black dots: primary star, Blue crosses: simulated secondary. Left: full data set (Obs1); Right: partial sample of the same dataset (Obs2). 
    Middle panels: CCF maps for both observations, each composed of combined CCFs for a range of $K_{\rm c}$ values, of the M-dwarf companion. Dashed lines and labels: Position of combined CCFs for $K_{\rm c}$, used in Fig.~\ref{fig:CCF_creation}. 
    Lower panels: Best fitting {\tt Saltire} models. Left, Fit for Obs1; Right, Fit for Obs2.}
    
    \label{fig:CCF_maps}.
\end{figure}
The orbital parameters of binaries from the EBLM project are typically very well determined from high resolution spectroscopy. Assuming two body motion, the companion shares all Keplerian orbital parameters, except the companion's argument of periastron ($\omega_{\rm c}$) and semi-amplitude ($K_{\rm c}$). The parameter $\omega_{\rm c}$ can be derived from the known primary orbit parameters: $\omega_{\rm c}=\omega - \pi$. The barycentric radial velocity of the companion will then take the form:
\begin{equation} \label{reflex}
    V_{r,c} = K_{\rm c} [\cos(\nu + \omega_{\rm c}) + e \cos(\omega_{\rm c})]
\end{equation}
With $\nu$ the true anomaly being determined by the period ($P$), the time of periastron ($T_{\rm 0,peri}$), the eccentricity ($e$) and well as $\omega$. 
$K_{\rm c}$, is not known at first, since it depends on the exact knowledge of the binaries mass ratio. For this reason CCF maps typically scan for a range of $K_{\rm c}$ which includes the expected semi-amplitude of the companion. To demonstrate this, we simulate 123 ESPRESSO \citep{pepe10} observations of the M-dwarf companion of the EBLM binary TOI-1338/BEBOP-1 \citep{Standing23} covering its full orbital period (Obs1 hereafter). A simulation has the advantage of avoiding to include the spectrum of the primary and the need to include any post processing in the analysis. The M-dwarf companion is simulated with PHOENIX model spectra \citep{Husser13} with $T_{\rm eff}=3300\,{\rm K}$, $\log\,g_\star=5.0$, and $\rm [Fe/H]=0.0$. We first correct wavelengths from vacuum to air, following equation 10 from \cite{Husser13}, then we match the spectral resolution to ESPRESSO ($R\sim140\,000$) using the implementation in the {\tt iSPec} package \citep{Blanco-Cuaresma14}, which convolves the spectra with a Gaussian kernel. 
Each spectrum is then shifted to match the companions reflex motion using Equation~\ref{reflex} as well as the primary's stellar and orbit parameters $P = 14.608558\,\rm d$, $T_{\rm 0,peri} = 2\,458\,206.16755\,\rm BJD$, $e=0.155522$, and $\omega = 2.05572\,{\rm rad}$ \citep{Standing23}, and the `true' semi-amplitude $K_{\rm c,true}= 77.84\,{\rm km\,s^{-1}}$, estimated from the mass ratio ($q = M_{\rm c} / M_1$) and the primary's semi-amplitude ($K_{\rm c,true} = K_{\rm 1} q$).
Finally, we cross-correlate the model spectra with an M2-dwarf line list, a type typically used for high-precision radial velocity measurements implemented by the ESPRESSO data reduction pipeline\footnote{The ESPRESSO pipeline has been publicly released on \hyperlink{https://www.eso.org/sci/software/pipelines/espresso/espresso-pipe-recipes.html}{https://www.eso.org}}. For the cross correlation, we convert the spectra to have a uniform sampling of $500\,{\rm m\,s^{-1}}$, which is close to the pixel resolution of the one dimensional ESPRESSO spectra and we sample the CCF for a velocity range between $-30\,{\rm km\,s^{-1}}$ and $30\,{\rm km\,s^{-1}}$ with a spacing of $1.5\,{\rm km\,s^{-1}}$.

Fig.~\ref{fig:CCF_creation} shows the CCFs for each data point when shifting the spectra for $K_{\rm c} = 76.7\,{\rm km\,s^{-1}}, 73.6\,{\rm km\,s^{-1}} {\rm, and}\,\,64.3\,{\rm km\,s^{-1}}$ respectively. The CCFs are less aligned to the companion's rest velocity the further $K_{\rm c}$ differs from $K_{\rm c,true}$. 
When we average each CCF to a combined CCF, a decrease of the combined CCF contrast as well as an asymmetry in the profile are clearly noticeable. By design, the width and shape of the combined CCF is defined by the shape of the single CCFs, as well as the average velocity shift of all spectra. This average velocity shift depends on the orbital parameters as well as of the phases, the orbit is sampled. For a well sampled circular orbit, the width is directly correlated to $K_{\rm c}$ - $K_{\rm c,true}$. The asymmetry of the profile is thus primary caused by the small orbital eccentricity of the simulated binary.

We now produce a CCF map and show it in Fig.~\ref{fig:CCF_maps}. This map represents the combined CCF functions for a range of $K_{\rm c}$ between $55 - 100\,\rm km\,s^{-1}$ calculated in steps of $1.5\,\rm km\,s^{-1}$. The position of the three combined CCFs, displayed in Fig.~\ref{fig:CCF_creation}, are now part of this map and their positions are highlighted for clarity. The resulting CCF map takes a cross-like shape, a saltire. This process resembles to what happens when optical telescopes are being focused. Far from the solution, the signal is defined and gets 'focused' at the correct semi-amplitude and rest velocity. We thus refer to the process of averaging the CCF signal for a certain semi-amplitude as K-focusing. 

Similar implementations have been described in literature for more than a decade now. For example \cite{Lafarga2023} describe their 'fast' process, which involves the cross-correlation in the telluric rest frame first, and combining the resulting CCFs after moving them into the companion's rest frame, using a certain semi-amplitude. Independent from the implementation, K-focusing is always the underlying process to create CCF maps in the $K_{\rm c}$ - $V_{\rm rest}$ plane.

On the right hand side panel of Fig.~\ref{fig:CCF_maps} we show how the shape of the CCF map changes when using a sub-set of 15 observations out of the 123 from the initial simulation that still cover the binary's orbit well (Obs2 hereafter). 

As described above, for real observations our assumption of identical contribution does not hold. We thus need to take into account a weighting of the contribution of each observation to the CCF map. This is typically the RMS noise of each spectrum,  but could also include for example phase-curve information of a specific molecule. Later in the paper (Sec.~\ref{planet}) we use actual observations where this effect is accounted for. 

\section{The \textit{Saltire} Model}

We introduce {\tt Saltire}\footnote{Saltire python code and documentation is available on \href{https://github.com/dsagred/saltire}{Github}. }, a simple, and scalable model to fit CCF maps, created from High Resolution Cross-Correlation Spectroscopy observations. For this model, we assume we can analytically express the CCF of the companion's contribution in velocity space at each time of the orbit. For precise radial velocity measurements the CCF is most often fitted with a simple Gaussian function, which works well for solar-type stars \citep{baranne96,pepe02}. For M-dwarfs, side-lobes are observed for reasons that are not completely understood and the simple Gaussian assumption does not hold. \cite{Bourrier18} showed that a double Gaussian fit can improve the CCF fitting of M-dwarf spectra. Side-lobes have also been observed for CCF signals of exoplanet atmospheres \cite[See e.g. Fig.8 in ][]{Brogi16}. Furthermore, we show in \cite{Sebastian23a} that data post processing methods, such as singular value decomposition can introduce a stretching of the signal which also produces artificial side-lobes on the CCF function. We thus simulate the CCF signal as a function of the relative rest velocity using a double Gaussian defined as:
\begin{equation} \label{bdgauss}
f(v) = h + ({A_1\mathcal{N}_1(\mu,\sigma_1) + A_2\mathcal{N}_2(\mu,\sigma_2)});
\end{equation}
with the mean height of the CCF outside the signal $ h$, $\mathcal{N}_1$ and $\mathcal{N}_2$ two Gaussian functions with respective intensities $A_1,\,A_2$ and standard deviations $\sigma_1,\,\sigma_2$ at identical velocity ($\mu $). To create the side-lobes $A_1$ and $A_2$ must have opposite signs. By fixing the velocity shift between the core (1) and the lobe component (2), we  create an asymptotic problem. When both amplitudes are allowed to increase, they would maintain a similar differential amplitude. This causes an increase of the calculation time for a least-squares minimisation algorithm. We avoid this by parametrising both amplitudes as:  
\begin{equation}
    A_1 = \frac{\Sigma}{(\Delta +1)};\,\,
    A_2 = \begin{cases} \rm
    0,& \text{if } \rm \Delta =  0\\
    \frac{\Sigma}{\left(\frac{1}{\Delta} +1\right)},              & \text{otherwise}
    \end{cases}
\end{equation}
By introducing the quotient $\Delta = A_2 / A_1$, as well as the sum $\Sigma = A_1 + A_2$ of both intensities. The parameter $\Sigma$ now represents the effective contrast of the CCF, which is independent from the amplitude of the side-lobes. It is negative for expected absorption and positive for emission spectra. We define $\Delta$ within the range $-1 < {\Delta} \leq 0$, which ensures that both amplitudes have an opposite sign. While $\Delta$ = 0, will basically turn the function into a single Gaussian, $\Delta\sim-1$ would describe the asymptotic case with both amplitudes being very large. By setting priors for $\Delta$, we can thus improve the convergence of the fit. 

Similar to Fig.~\ref{fig:CCF_creation}, we can now derive the CCF signal $f(v)$ for each observation time. The relative rest velocity of the CCF signal can be derived from the known Keplerian orbital parameters and Equation~\ref{reflex} as, 
\begin{equation}
v = V_{r,c}(K_{\rm c}) - V_{r,c}(K_{\rm c, true}) + V_{\rm rest},
\end{equation}
for a given $K_{\rm c}$. This adds two free parameters to our model, the true semi-amplitude ($K_{\rm c, true}$) and the rest velocity ($\rm V_{\rm rest}$) of the detection.
Combining the CCF functions $f(v)$ for all observation times, we can derive a combined CCF function $f'(K_{\rm c})$ for each sampled value of $K_{\rm c}$. This process is identical to the K-focusing, described in Sec.\,\ref{model}. Therefore, we can combine all sampled $f'(K_{\rm c})$ into a model CCF map. This step is by design identically to what we described for simulated combined CCFs in Sec.\,\ref{model}. This method assumes that all RV trends e.g. due to a third body in a wider orbit have been removed prior the analysis. A good understanding of the primary star's radial velocity trends is typically the case for high-precision spectroscopy of binaries or planet host stars.

Our definition of $V_{\rm rest}$ as the rest velocity of the detection means that the systemic velocity ($V_{\rm sys}$) has not been removed. Since it is a free parameter in this model, it can simply be applied to the often used velocity definition with $V_{\rm sys}$ being removed prior the measurement.

\subsection{Fitting CCF maps made easy}

The {\tt Saltire} model can be used to fit CCF maps. We implement the model as a residual function using the {\tt lmfit} framework to apply least-squares fitting \citep{Newville16}. In this section we summarise the basic usage of the model and input parameters. 
Due to the model construction, the only free parameters are $K_{\rm c}$, $V_{\rm rest}$, as well as the parameters of the one dimensional function $f(v)$, which are $h,\, \Sigma,\, \Delta,\,\sigma_1,$ and $\sigma_2$. Fixed parameters are simply identical to the parameters which have been used to create the CCF map of the simulated data described earlier (Obs1 and Obs2). Here we divide three sets of input parameters: (i) the CCF map dimensions, which are the sampled velocity and semi-amplitude ranges $[V_{\rm rest}]$, and $[K_{\rm c}]$, (ii) the observation contribution parameters ({\tt obs}), which are observation times and weights for each observation, and (iii) the fixed orbit parameters ({\tt fixpar}), which include the period ($P$), time of periastron ($T_{\rm 0,peri}$), eccentricity ($e$), and argument of periastron ($\omega$). Latter are assumed to be arrays with length of the semi-amplitude range $[K_{\rm c}]$, to allow fitting of possible different orbit parameters for each sampled semi-amplitude.
Additionally, {\tt Saltire} has a built-in ability to replace the double Gaussian function by a double Lorentzian function. Due to similarities between the parameters of both functions, the change can be done by simply setting the {\tt func} keyword.

To apply priors, we make use of the {\tt lmfit} {\tt param} object. For $\Delta$ the maximal value is zero (no side-lobes) and the minimal value should be close, but $> -0.5$ (strong side-lobes) to avoid the fit entering the asymptotic space. We also ensure the expected width of the side-lobes $\sigma_2$ is always larger than the line width $\sigma_1$. Despite these recommendations, all free parameters and their limits can be freely selected prior to the fit. 

The fitting is implemented by using the {\tt saltirise} function. It returns the residuals normalised by the CCF errors $\rm (data-model)/error$ as a flattened array for each combination of $V_{\rm rest}$, and $K_{\rm c}$, or the initial model, if no data are provided. Despite, the least-squares fit as implemented in {\tt lmfit} allows to derive correlations and error propagation for each free parameters, {\tt Saltire} also implements a {\tt run\_MCMC} function. This is to sample the posterior probability distribution of the {\tt Saltire} free model parameters, using the Markov chain Monte Carlo (MCMC) code \texttt{emcee} \citep{Foreman-Mackey13}. With the MCMC, an extra parameter is fit, the CCF-jitter ($\sigma_{\rm jit}$) of the data that optimally weights the fit in the log-likelihood function. The jitter is a way to avoid under-estimated uncertainties, and accounts for RMS noise, as well as for correlated signals like spurious CCF correlations of stellar lines and/or post processing artefacts. Uniform priors can be set using the same {\tt param} object described above, but adding $\sigma_{\rm jit}$ as additional parameter. 
%

\begin{figure*}
    
    \includegraphics[width=\linewidth]{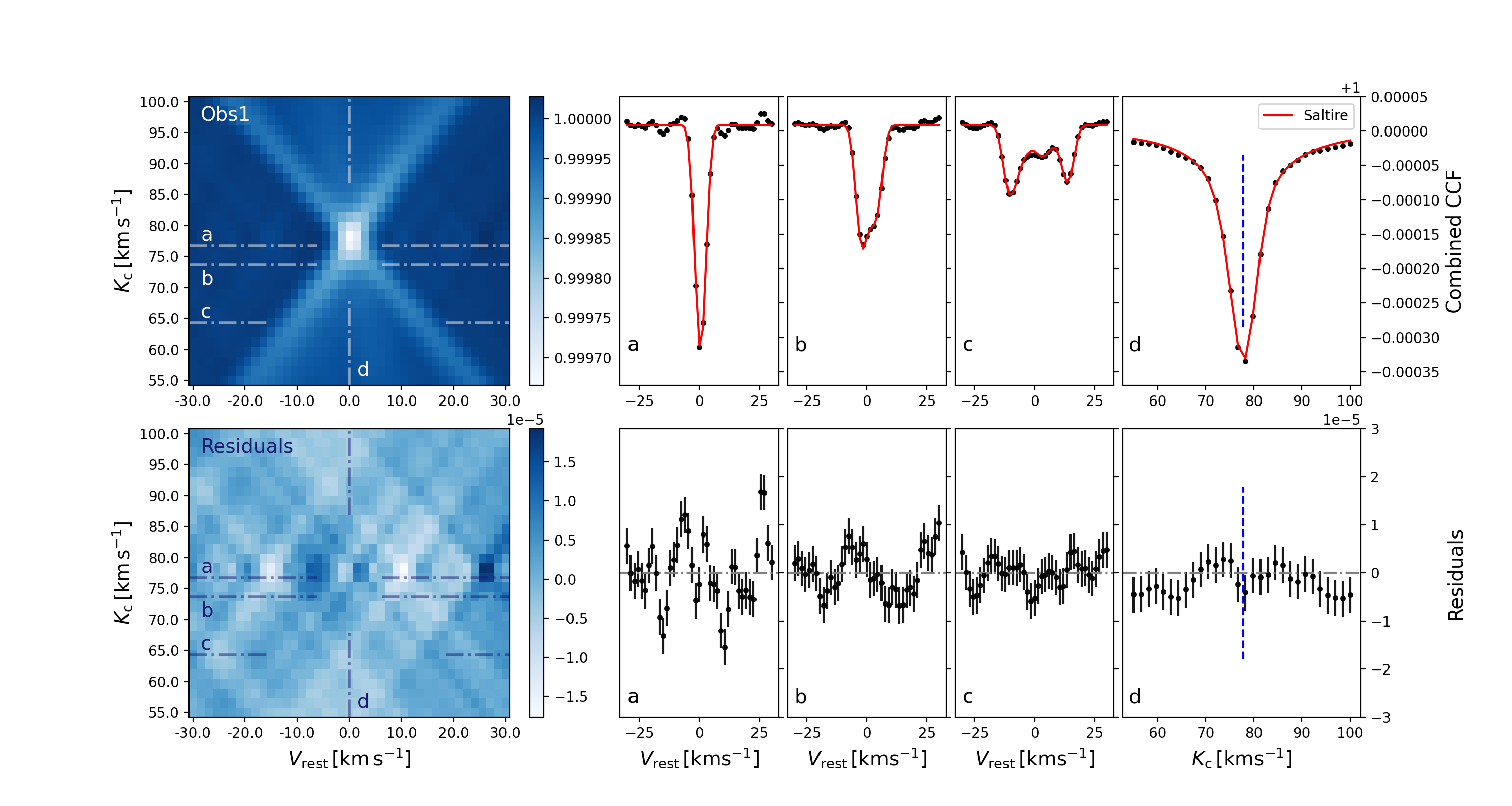}
    \caption{Upper panel; Left: CCF map for Obs1. Dashed lines and labels: Positions of cuts though the data, shown on the right. Right: Cuts through the data, for different semi-amplitudes (a,b,c), and for the rest velocity of maximum CCF contrast (d); Black dots: data for simulated data-set Obs1, Red line: best fitting {\tt Saltire} model. Green dashed line: mark of the input semi-amplitude of the EBLM companion.
    Lower panel; Left: Residual CCF map after removing the {\tt Saltire} model. Dashed lines and labels: Positions of cuts though the residuals, shown on the right. Right: Cuts through the residuals at similar position as above. Error bars represent uncertainties, derived from 2D {\tt Saltire} fit to the the CCF map.}
    \label{fig:CCF_cuts}
\end{figure*}

\begin{figure*}
    
    \includegraphics[width=\linewidth]{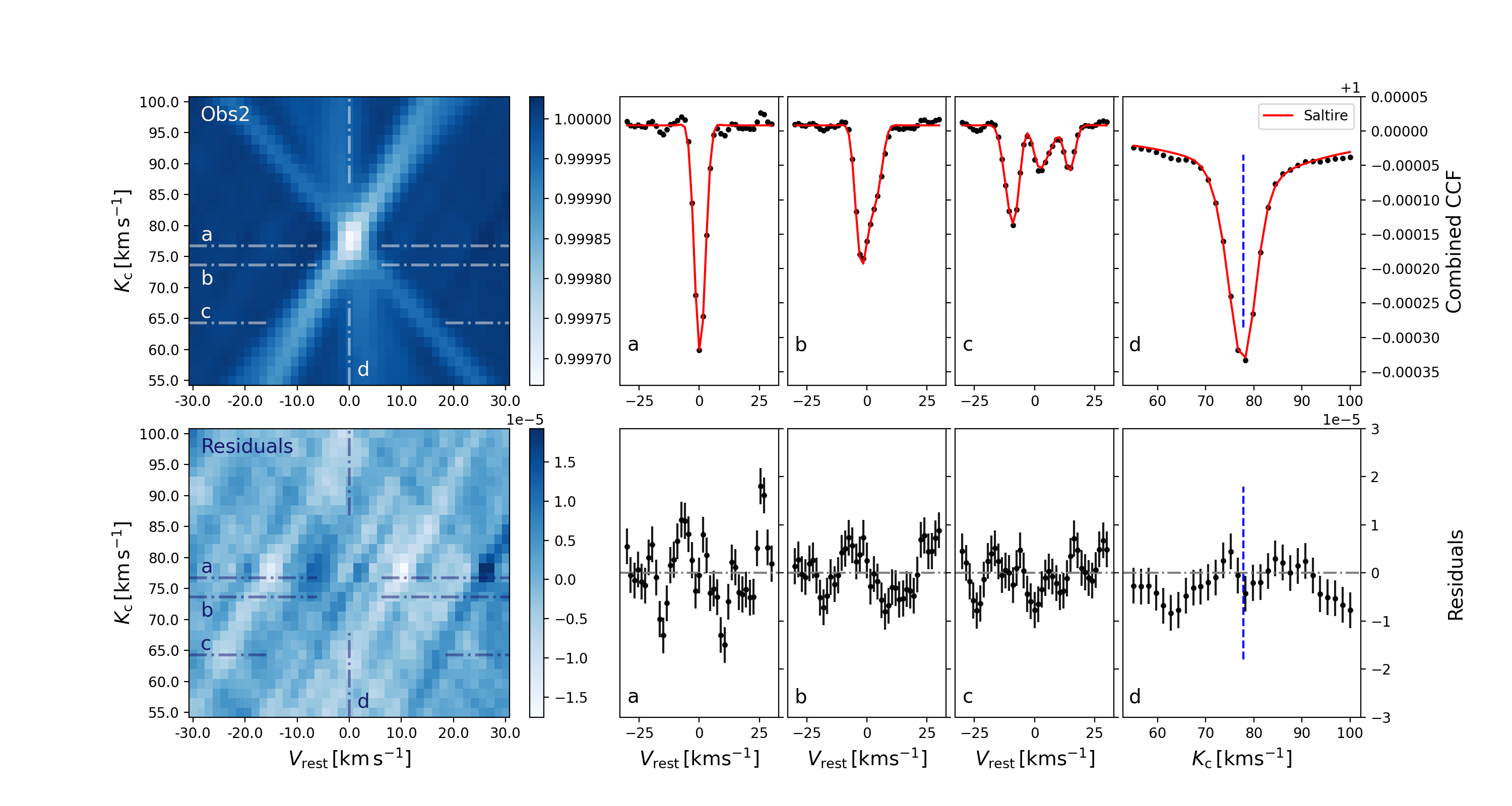}
    \caption{Upper panel; Left: CCF map for Obs2. Dashed lines and labels: Positions of cuts though the data, shown on the right. Right: Cuts through the data, for different semi-amplitudes (a,b,c), and for the rest velocity of maximum CCF contrast (d); Black dots: data for simulated data-set Obs2, Red line: best fitting {\tt Saltire} model. Green dashed line: mark of the input semi-amplitude of the EBLM companion.
    Lower panel; Left: Residual CCF map after removing the {\tt Saltire} model. Dashed lines and labels: Positions of cuts though the residuals, shown on the right. Right: Cuts through the residuals at similar position as above. Error bars represent uncertainties, derived from 2D {\tt Saltire} fit to the the CCF map.}
    \label{fig:CCF_cuts2}
\end{figure*}

\begin{table*}
	\centering
	\caption{{\tt Saltire} Fit results for both simulated EBLM data-sets.}
	\label{tab:fit_model}
	\begin{tabular}{lccccc}
    \hline
    Parameters & \multicolumn{2}{c}{\textit{Obs1}}              & \multicolumn{2}{c}{\textit{Obs2}} & Input fit parameter\\
                    & Least-squares         & MCMC              & Least-squares     & MCMC &\\\hline
    $K_{\rm c}\,[\rm km\,s^{-1}]$     & 77.841$\pm$0.011      & 77.840$\pm$0.012  & 77.828$\pm$0.013  & 77.828$\pm$0.013  & 77.840 (inserted)\\
    $V_{\rm rest}\,[\rm km\,s^{-1}]$  & 0.4691$\pm$0.0088     & 0.4690$\pm$0.0083 & 0.4448$\pm$0.0092 & 0.4446$\pm$0.0090   & 0.0 (inserted)\\
    $\Sigma$ & (-3.506$\pm$0.016)e-04& (-3.510$\pm$0.015)e-04 & (-3.504$\pm$0.014)e-04 & (-3.507$\pm$0.013)e-04 & -3e-04\\
    $\Delta$& -0.42$\pm$2.31          & -0.26$\pm$0.14  & -0.49$\pm$1.04    & -0.31$\pm$0.13    & -5e-3\\
    $\sigma_1\,[\rm km\,s^{-1}]$    & 2.32$\pm$0.56         & 2.267$\pm$0.055   & 2.38$\pm$0.28   & 2.314$\pm$0.049   & 2.4\\
    $\sigma_2\,[\rm km\,s^{-1}]$   & 2.6$\pm$1.4         & 2.71$\pm$0.18       & 2.66$\pm$0.60     & 2.78$\pm$0.15     & 4.58\\
    $h$        & 1+(8.52$\pm$0.16)e-6  & 1+(8.53$\pm$0.15)e-6 & 1+(8.31$\pm$0.17)e-6 & 1+(8.23$\pm$0.17)e-6 & 1.0\\
    $\sigma_{\rm jit}$    & -- & (3.729$\pm$0.079)e-06 & -- &  (4.155$\pm$0.083)e-06 & --\\      
    \hline
    \end{tabular}
    \\
\end{table*}

\section{Application to simulated HRCCS EBLM data} \label{binary}

We estimate how well the model fits CCF maps by using two simulated data-sets of a high-contrast binary, as described in Sec.\,\ref{model}, Obs1 and Obs2. We first do a least-squares minimisation, followed by an MCMC sampling. As starting parameters, we use input values for $K_{\rm c}$ and $V_{\rm rest}$ and CCF function parameters listed in Tab.\,\ref{tab:fit_model} with wide uniform priors. The least-squares results are used as starting parameters for the MCMC sampling, also using wide uniform priors. We run 42 parallel chains with 4\,000 samples each. The first 1\,500 samples of each walker are rejected (the burn-in) and the resulting samples are thinned by a factor 5, resulting in a posterior distribution of 21\,000 independent samples for each parameter. The reported parameters are derived as the 50th percentile, and errors by averaging the 15.8655/84.1345 percentiles of the posterior distribution.

Fig.\,\ref{fig:CCF_maps} shows {\tt Saltire} models generated for both data-sets. By its very design, the model computes different CCF map shapes corresponding to different data sampling and phase coverage. Tab.\,\ref{tab:fit_model} shows the results of the CCF map fit with {\tt Saltire} for both observations. The resulting MCMC samples for both sets of simulated observations are displayed in Fig\,\ref{fig:Obs1_MCMC}, and Fig\,\ref{fig:Obs2_MCMC}.  

Fig.\,\ref{fig:CCF_cuts} reveals the shape of the function by performing cuts through the CCF maps as well as the best fitting {\tt Saltire} models. Cuts for constant semi-amplitudes are shown for  (a) $K_{\rm c} = 76.7\,{\rm km\,s^{-1}}$, (b) $73.6\,{\rm km\,s^{-1}}$, and (c) $64.3\,{\rm km\,s^{-1}}$. Since we use synthetic spectra, no side-lobes are generated. Due to the absence of side-lobes the parameters $\Delta$ and $\sigma_2$ are not well constrained. We find that repeating the fit with both parameters fixed, does not change the result for the other parameters and uncertainties. A cut is shown for constant rest velocity close to the maximum CCF contrast (d) $V_{\rm rest}=0\,{\rm km\,s^{-1}}$. This step is used in most studies to determine $K_{\rm c}$. Fig.\,\ref{fig:CCF_cuts2} shows the same for Obs2, highlighting the shape differences for both samples, which can best be seen for cuts b and c, hence for larger deviations from $K_{\rm c, true}$.

The measurements for $K_{\rm c}$ returned by the model agree with the inserted value $K_{\rm c, true}$ for the simulated observations, within $\rm 1\sigma$. Despite being in statistical agreement, we note that there is a small difference of $-12\,\rm m\,s^{-1}$ between Obs1 and Obs2. We also find that the values returned for $V_{\rm rest}$ by the model show a difference of $24\,\rm m\,s^{-1}$ ($2.6\,\rm \sigma$). The origin of these discrepancies is discussed in Sec.~\ref{wiggles}. We note that the measured $V_{\rm rest}$ shows a similar offset of about $450\,\rm m\,s^{-1}$ for both Obs1 and Obs2. This offset is not surprising, since (i) the model spectra have been corrected from vacuum wavelengths, and (ii) the M-dwarf line list has been compiled using observed spectra. Both effects can be credited for this systematic offset. 

To test the influence of the shape difference between Obs1 and Obs2, we used a simple one-dimensional Gaussian fit to measure $K_{\rm c}$ for the rest velocity of maximum CCF contrast (cut d). This test also simulates what is traditional employed within the literature to measure $K_{\rm c}$ from CCF maps. The best fitting value for $K_{\rm c}$, we find this way is $77.786\pm0.088\,\rm km\,s^{-1}$ and $77.694\pm0.069\,\rm km\,s^{-1}$ for Obs1 and Obs2, respectively. These measurements are 54\,$\rm m\,s^{-1}$ (0.6 $\rm \sigma$) and 146\,$\rm m\,s^{-1}$ (2.1$\rm \sigma$) from the inserted value $K_{\rm c, true}$ respectively. Thus, we can confirm that the shape difference between both data-sets results in a systematic deviation, even at maximum CCF contrast, which - if not modelled properly -  will result in discrepant measurements which can be on the order of 10 times larger, compared to what we can achieve with the current implementation of {\tt Saltire}.

\subsection{K-Focusing of CCF wiggles} \label{wiggles}

Since the simulated observations are by definition noise-less we ought to retrieve exact values, for both observations, but we do not. Therefore, the differences we retrieve must be created by the analysis. 
Spurious CCF correlations between the template and other parts of the spectrum are a prime suspect. They are a feature of CCFs \citep[nicknamed {\it wiggles},][]{Lalitha23}.

In real observations these wiggles are quasi stationary, thus show only small changes due to changing observing conditions. In our simulated observations, these are exactly identical, for each observation. When creating the CCF map, these wiggles undergo the same K-focusing process (see Sec.\,\ref{model}), than the main CCF signal. We can now use these simulated, observations to study the effects, introduced by wiggles. The simulated stationary wiggles can be seen in the residuals panels of Fig\,\ref{fig:CCF_cuts} for Obs1 and Fig\,\ref{fig:CCF_cuts2} for Obs2.
They show a decreasing amplitude for larger deviations from $K_{\rm c,true}$ (from a to c). This is expected from the K-focusing process, which basically causes a smearing by averaging them for increasing RV shifts. We also note that the overall shape of the wiggles for Obs1 and Obs2 appear very similar. This is caused by the fact that the orbit parameters and selected range for $K_{\rm c}$ are identical for both observations.

\begin{figure}
	
	\includegraphics[width=\linewidth]{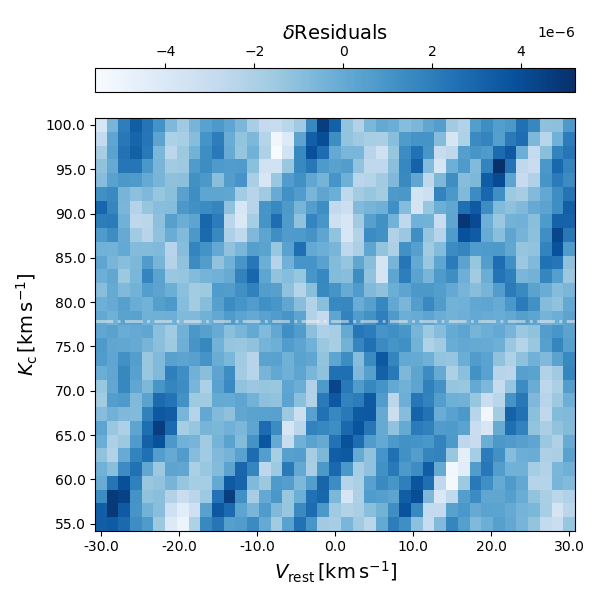}
    
    \caption{Difference between the residual maps from Obs1 and Obs1. Remaining structures are mainly caused by residual wiggles due to the phase differences between both observations. Dashed line: position of inserted semi-amplitude ($K_{\rm c,true}$).
    }
    \label{fig:Residual_dif}
\end{figure}

A crucial aspect is that, the shape - originated by the K-focusing process - is also dependent on the phase coverage. Since both observations have a different phase coverage - similar to the companions CCF shape - the wiggles must be different for both observations. Fig.~\ref{fig:Residual_dif} shows the difference between the residual maps of Obs1 and Obs2 with the inserted value of $K_{\rm c,true}$ highlighted. Wiggles of the CCF map are identical at $K_{\rm c,true}$, for both observations, thus are largely removed in this position, but differ increasingly - due to K-focusing - with larger semi-amplitude deviations. Fig.~\ref{fig:Residual_dif} thus shows that residual correlations caused by the wiggles must be the dominating structure.

This explains why we measure slightly different values for $K_{\rm c}$ and $V_{\rm rest}$: The differences shown in Fig.~\ref{fig:Residual_dif} are currently not modelled by {\tt Saltire}.
Nevertheless, we show, how 1D wiggles at $K_{\rm c,true}$ propagate in the 2D CCF map due to the K-focusing process. Modelling the 2D wiggles is not the goal of this paper, but future applications to the {\tt Saltire} model could include a model of these quasi stationary wiggles into the CCF function $f(v)$ and thus use them as an analytic tool to disentangle between the CCF signal and other aliases from post processing with the goal to measure yet more accurate parameters of the companion, using CCF maps.

\section{Application to exoplanet HRCCS observations} \label{planet}

Providing a framework to accurately measure atomic species and molecules in exoplanet atmospheres will be one of the main use-cases for {\tt Saltire}. We, show its application to the close-in gas giant planet $\rm \tau$ Bo\"otis b. \cite{Brogi12} (B12 hereafter) reported the detection of carbon monoxide (CO) in the atmosphere of this planet by observing the day side emission of the planet. They used CRIRES \citep{Kaeufl04} observations with a wavelength range of $2.287-2.345\,\rm \mu m$ and a resolution of $R~100\,000$ to monitor the planet during three nights. These data cover the orbital phase before, close to, and after the phases of superior conjunction. The data were post-processed identically as in B12, to clean telluric and stellar lines. The orbital radial velocity parameters have been presented is several studies \citep[e.g.][]{Butler06,Brogi12,Borsa15,Justesen19}. These also report a radial velocity trend of the F6V host star, due to a wide M-dwarf companion. We use the circular orbital solution presented by B12 for this demonstration as fixed parameters ($P =3.312433\,\rm d$, $T_{0,\rm peri} = 2\,455\,652.108\,\rm HJD$). In here we neglect the expected trend, which is less than $1\,\rm m\,s^{-1}$ and, thus below the velocity resolution of CRIRES during to the short observing period for the B12 data. Similarly to Sec.~\ref{binary} we cross-correlate the post processed spectra with a line mask. The line-mask is generated using HITRAN \citep{HITRAN2016} line positions for $\rm ^{12}C^{16}O$, which were weighted by the CO model, described by B12. Using a CO line mask allows us to derive the mean line profile of the exoplanet atmosphere without prior assumption of its shape. We sample $K_{\rm c}$ from 50 to $170\,\rm \,km\,s^{-1}$ in steps of $1.5\,\rm km\,s^{-1}$, and $V_{\rm rest}$ from -70 to 50$\rm \,km\,s^{-1}$ in steps of $1.5\,\rm km\,s^{-1}$. Our selected step width corresponds to the pixel resolution of the CRIRES data. Both ranges are selected to include the position of the CO detection, reported in B12 ($K_{\rm c} = 110\pm3.2\,{\rm km\,s^{-1}}$) as well as the systemic velocity of the host star reported in \cite{Donati08} ($V_{\rm sys} = -16.4\pm0.1\,{\rm km\,s^{-1}}$)\footnote{In B12, this systemic velocity was used as fixed input parameter for $V_{\rm rest}$ to optimise the orbital phase shift.}. We co-add the CCF maps from different CRIRES detectors similarly to B12 by weighting them by the relative line intensities of the CO lines in each detector.

\begin{figure}
	
	\includegraphics[width=\linewidth]{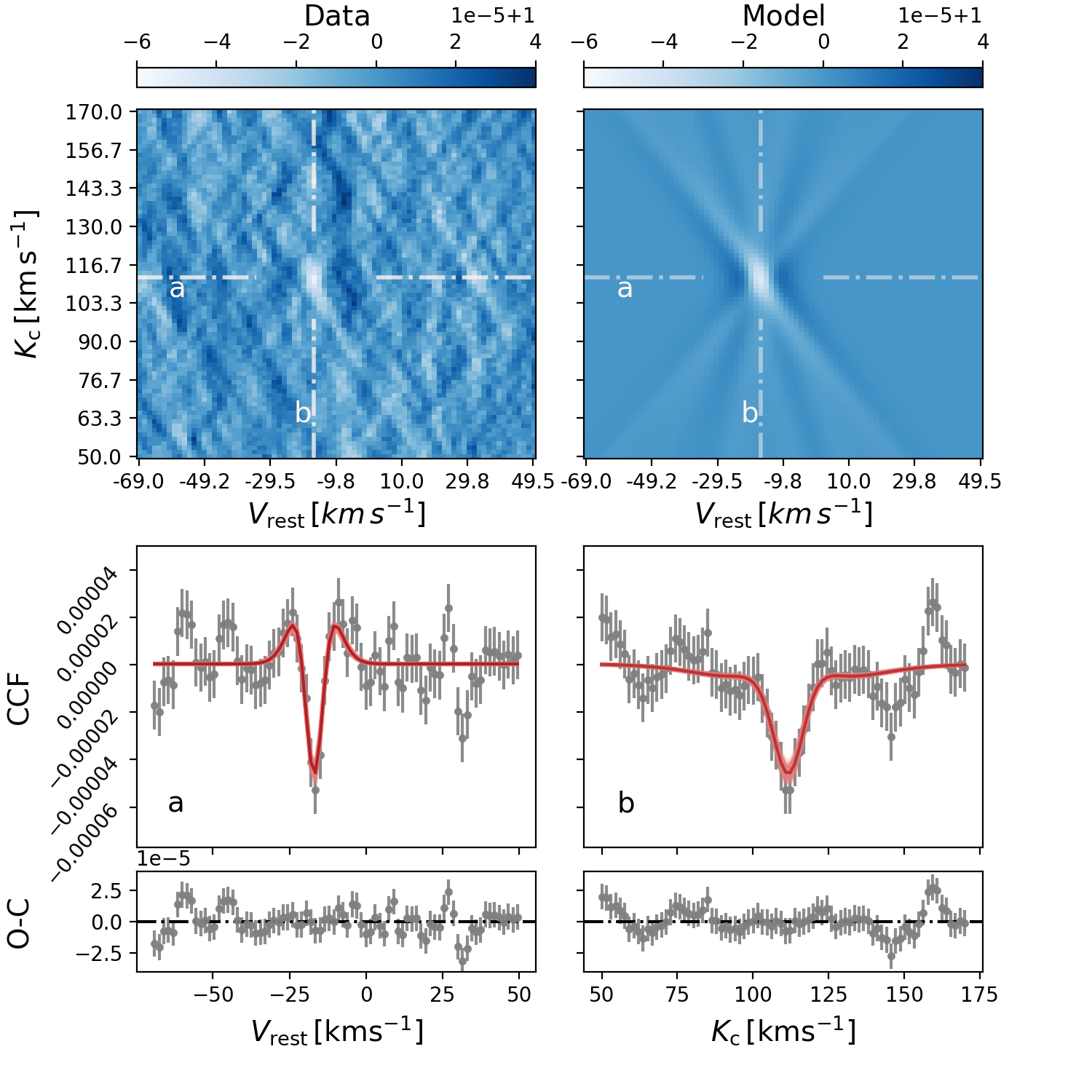}
    
    \caption{
    Upper panel, left: CO CCF-map of $\rm \tau$ Bo\"otis b, reproduced from \citet{Brogi12} data, right: {\tt Saltire} best fitting model. Dashed lines indicate the position of maximum CCF contrast of the combined CCF map. The line orientation indicates the cuts - shown in the lower panels - for clarity; Lower panel: Cuts trough the CCF map at maximum CCF contrast for both axis. Grey dots: CCF-map data and residual data (lower panel). Error-bars represent the jitter parameter, derived from MCMC sampling; Red line: best {\tt Saltire} model fit; Red shaded lines: {\tt Saltire} models from posterior samples.
    }
    \label{fig:TauBoo_fit}
\end{figure}

We obtain the best fitting {\tt Saltire} model to the combined CCF-map by least-squares minimisation using the input parameters given in Tab.~\ref{tab:fit_TauBoo}. The resulting fit parameters are used as starting parameters to sample the posterior distribution using the MCMC with the same number of walkers and samples,  as described in Sec.~\ref{binary}. The 50th percentile, and average errors from the posterior distributions are presented in Tab.~\ref{tab:fit_TauBoo}. For the parameters $K_{\rm c}$ and $V_{\rm rest}$, we additionally report systematic uncertainties in excess of the MCMC errors (see Sec.~\ref{uncertainties} for a detailed discussion).

Fig.~\ref{fig:TauBoo_fit} shows the resulting CCF map for the combined observations as well as the best fitting {\tt Saltire} model. We can use the best fitting parameters to estimate the significance of the detection as the CCF contrast ($\Sigma$) divided by the jitter term ($\sigma_{\rm jit}$). We find that the CO signal is resolved with a SNR of $4.6\,{\rm \sigma}$ at the expected position. With {\tt Saltire} we can precisely model the expected shape of this CO detection. It shows three cross-like structures, which are mostly buried in the noise for these data. The lower panel of Fig.~\ref{fig:TauBoo_fit} shows a cut of the combined {\tt Saltire} model at maximum contrast (a). The double Gaussian structure of our best fitting model function is well fitted, due to the side-lobe like structure of the data. The posteriors of the MCMC parameter for the width of the side lobes ($\sigma_2$) is well defined (visualised in Fig.~\ref{fig:TauBoo_MCMC}). The parameter $\Delta$ shows a trend towards -0.5. Due to the asymptotic nature of the double Gaussian function, increasing the prior towards -1 will still result in the lowest value being favoured, as soon as side lobes are fitted. Since this will by design only allow marginal changes of the fit close to the numeric precision level, we keep the lower bound of this prior at -0.49.

The cuts demonstrate that the assumption of a Gaussian shape also holds for these exoplanet observations. We show the best fit for a cut trough the semi-major axis close to the maximum CO detection (b) $V_{\rm rest}=-16.5\,{\rm km\,s^{-1}}$. 

Our {\tt Saltire} measurement for $K_{\rm c}$ and $V_{\rm rest}$ are statistically identical to the result reported by B12 (within $1\,\rm \sigma$). This is not surprising, since we used the same data, as well as the same orbital parameters, that have been derived by B12 to match the systemic velocity of the host star. Small differences below the uncertainties are not surprising as well, given our use of a different method for the cross-correlation, which is based on a line mask and not on a model spectrum.

Similarly to Sec.~\ref{binary}, we fit a one-dimensional Gaussian model to the rest velocity of maximum CCF contrast which results in $K_{\rm c} = 111.29\pm0.58\,{\rm km\,s^{-1}}$. The one dimensional Gaussian fit differs by about $540\,{\rm m\,s^{-1}}$ from the {\tt Saltire} fit (being statistically in agreement, when adding systematic uncertainties of $2-3\,{\rm km\,s^{-1}}$, see Sec.~\ref{uncertainties}). 

\begin{table}
	\centering
	\caption{{\tt Saltire} Fit results for atmospheric CO detection of $\rm \tau$ Bo\"otis b. The uncertainties for the parameters $K_{\rm c}$ and $V_{\rm rest}$ are a combination of MCMC fit errors and systematic uncertainties.}
	\label{tab:fit_TauBoo}
	\begin{tabular}{lcc}
    \hline
    Parameters & MCMC results & Input fit parameters\\\hline
    $K_{\rm c}\,[\rm km\,s^{-1}]$ & 111.8$\pm$2.4 (0.27, 2.39)$^{**}$& 110$^{*}$\\
    $V_{\rm rest}\,[\rm km\,s^{-1}]$  & -16.86$\pm$0.50 (0.12, 0.48)$^{**}$ & -16.4$^{*}$\\
    $\Sigma$ & (-4.70$\pm$0.22)e-5 & -4e-05\\
    $\Delta$& -0.468$\pm$0.024& -0.1\\
    $\sigma_1\,[\rm km\,s^{-1}]$ & 2.85$\pm$0.11 & 1.5\\
    $\sigma_2\,[\rm km\,s^{-1}]$ & 5.92$\pm$0.25 & 3.5\\
    $h$        & 1+(3.1$\pm$1.7)e-7& 1.0\\
    $\sigma_{\rm jit}$ &  (1.022$\pm$0.0088)e-5 & --\\
    \hline
    \multicolumn{3}{l}{$^{*}$ \cite{Brogi12}}\\
    \multicolumn{3}{l}{$^{**}$ Error components: (fit error, systematic uncertainty).} \\
    \end{tabular}
    \\
\end{table}

\subsection{Uncertainties - Model Robustness vs. correlated noise} \label{uncertainties}

The $V_{\rm rest}$ and $K_{\rm c}$ sampling of the CCF map might be crucial when estimating fit errors. The uncertainty determination of the least-squares fit, as well as the jitter term derived by the MCMC, relies on the assumption that the uncertainties of adjacent data points are uncorrelated with each other. Violating the assumption of uncorrelated data will result in underestimated uncertainties from the fit.

For the $V_{\rm rest}$ range, this can easily be achieved by selecting a sampling that is wider than the average pixel resolution of the instrument. In this case each data point originates from a superposition of different pixels of the detector full-filling the requirement of uncorrelated data. 

The $K_{\rm c}$ range is more difficult. As noted by \cite{Hoeijmakers2020}, data in the $K_{\rm c}$ range are partly correlated, affecting the error determination using noise statistics. This correlation is a result of the K-focusing process. Each adjacent data point in $K_{\rm c}$ direction is a superposition of the same spectra. Those have been shifted following the reflex motion, at a different semi-amplitude, which is defined by the $K_{\rm c}$ spacing. For close-in exoplanets with circular or low-eccentric orbits for instance, the maximum RV-shift is close to the $K_{\rm c}$ spacing itself, and minimum RV-shifts are close to zero. For a closely sampled orbit, or observations during a transit/eclipse, some spectra, and sometimes all of them, are only shifted by velocities below the pixel resolution. Notably, even white noise of the spectra will, thus, become part of the correlated signal in the CCF map. To study the influence of correlation on uncertainties and measured parameters, we derive the geometrical correlation of each data point in the CCF map as the fraction of combined spectra which have been RV-shifted by less than the instrumental resolution.

For the simulated high-contrast binary (Obs1, see Sec.~\ref{model}), a $K_{\rm c}$ sampling of $1.5\,{\rm km\,s^{-1}}$ - as we use in this study - results in 21\% correlation between each adjacent data point in the $K_{\rm c}$ range. For the CO detection of $\rm \tau$ Bo\"otis b, all adjacent data points are shifted below the pixel resolution (100\% correlation) - due to the observation at orbital phases close to superior conjunction.

\begin{figure}
	\includegraphics[width=\linewidth]{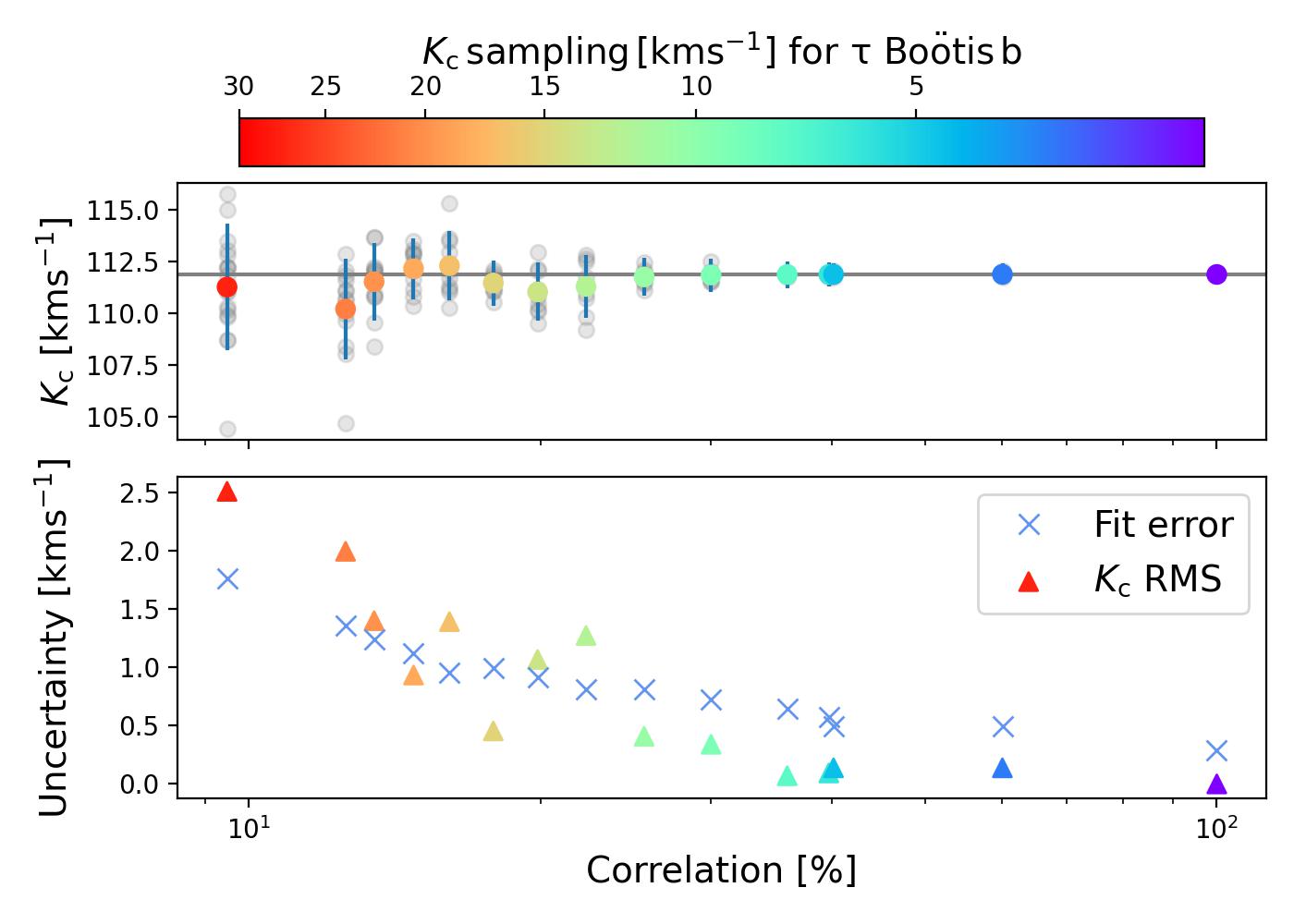}
    \includegraphics[width=\linewidth]{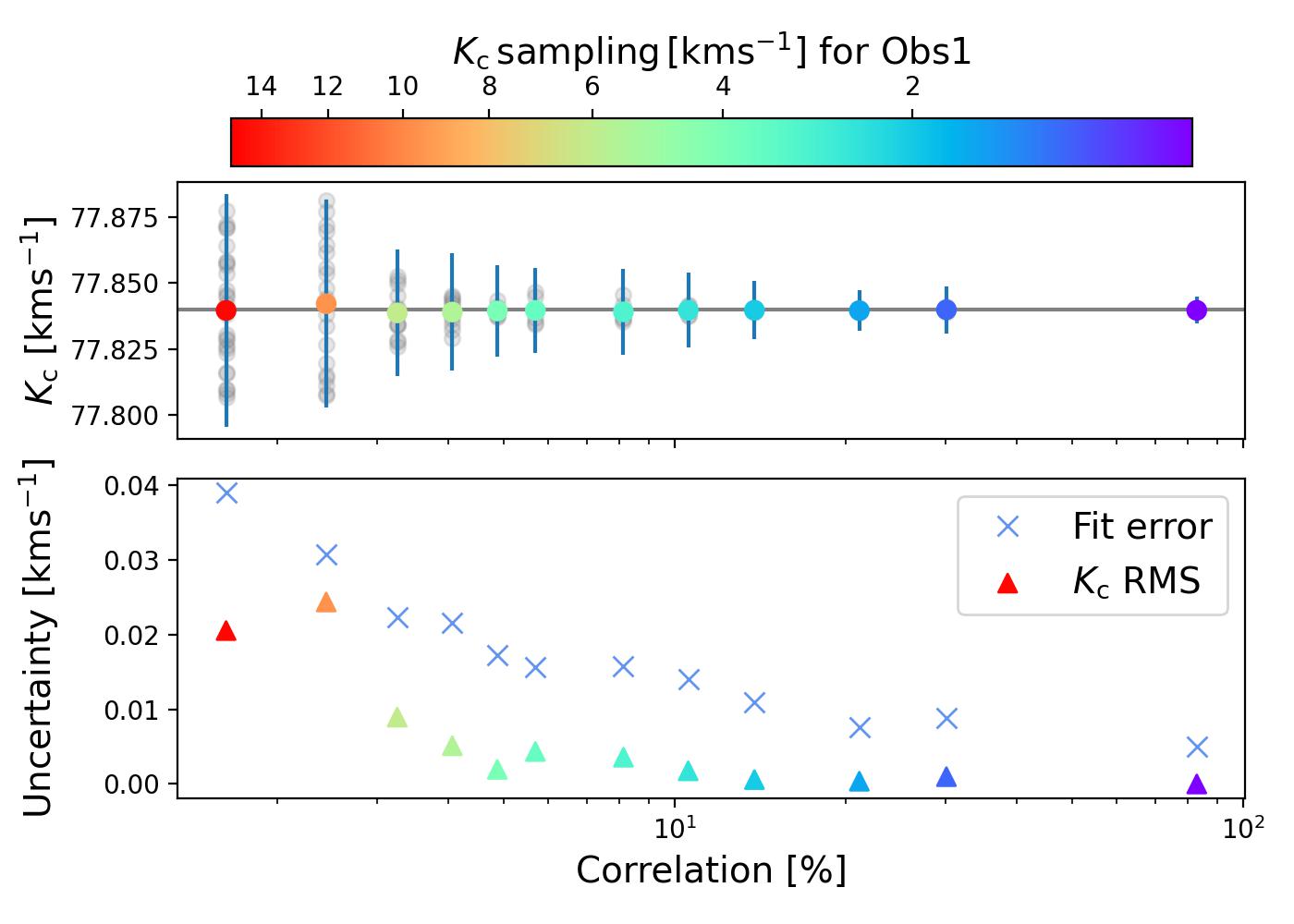}
    
    \caption{Uncertainty analysis from $K_{\rm c}$ spacing for $\rm \tau$ Bo\"otis b as well as for the simulated high-contrast binary (Obs1). Upper panel: Measurement of $K_{\rm c}$ from partial maps with different $K_{\rm c}$ spacing. Gray dots: Individual measurements for partial maps. Coloured dots: average $K_{\rm c}$ of individual measurements. Colour coding $K_{\rm c}$ sampling, of partial map.
    Lower panel: Components of uncertainties. Blue crosses: Average uncertainties from individual fits. Coloured triangles: RMS of individual $K_{\rm c}$ measurements. Colour coding as above panel.
    }
    \label{fig:sampling}
\end{figure}

We explore in the following two sections the effects of correlated data on the uncertainties when fitting CCF maps with {\tt Saltire}.

\subsubsection{$K_{\rm c}$ sampling}

At first we test for the impact of the $K_{\rm c}$ sampling on the $K_{\rm c}$ parameter, returned by {\tt Saltire}. For this, we split the CCF map evenly in $K_{\rm c}$ space. For instance, taking every second row of the CCF map will result in two partial CCF maps, each with half the $K_{\rm c}$ sampling, hence reducing the correlation between adjacent data points. The more maps we create, the less correlated are each data from one another. By analysing each of these partial CCF maps independently, we can test for systematics not taken into account by the fit of the original CCF map.

For each set of partial maps, we then fit the {\tt Saltire} model and extract $K_{\rm c}$ and its uncertainty using the least-squares fit. By doing so, we can compare the scatter of the returned $K_{\rm c}$ values, for similarly correlated data to the average fit error returned by {\tt Saltire}. For $\tau\,\,$Bo\"otis\,b data, we start from a CCF map close to the pixel resolution ($1.5\,{\rm km\,s^{-1}}$).

The upper panel of Fig.~\ref{fig:sampling} shows the dependence of the returned $K_{\rm c}$ values as well as the average $K_{\rm c}$ value for differently correlated maps. The average value shows an increasing uncertainty the less the CCF map is sampled in $K_{\rm c}$ range. We derive the uncertainty by adding quadratically the average of the individual fit errors to the standard deviation of the individual measurements of $K_{\rm c}$. The contributions of both components are shown in the lower panel of Fig.~\ref{fig:sampling}. Both uncertainties follow the same trend, which is governed by the data-set sampling. They become comparable for a $K_{\rm c}$ sampling larger than $10\,{\rm km\,s^{-1}}$ ($< 25\%$ correlation). We then repeat the same test for Obs1 of our simulated high-contrast binary, starting from a CCF map with a $K_{\rm c}$ sampling of $0.5\,{\rm km\,s^{-1}}$. In this noise-less observations the fit uncertainties, returned by {\tt Saltire} are always larger than the deviations (RMS) of the individual measurements. Also no residual deviations of $K_{\rm c}$ are observed.

Using this test, we do not find systematic uncertainties in obvious excess of the reported fit uncertainties. We note that this test relies on the sampling of the signal itself, resulting in a lower accuracy the more sparsely the signal is sampled in the $K_{\rm c}$ range. It is, thus, sensitive to detect systematics from correlated data that have a correlation length scale in $K_{\rm c}$ smaller than the signal itself.

We further show in the upper panel of Fig.~\ref{fig:sampling} that the average $K_{\rm c}$ parameter is largely invariant from the $K_{\rm c}$ sampling. {\tt Saltire} can therefore be used to derive accurate parameters even for sparsely sampled data sets, which might be an opportunity to decrease computation costs - which are usually involved - when deriving highly sampled CCF maps.

\subsubsection{Global correlated noise}

At second, we test the global impact of correlated noise on the measurement of $K_{\rm c}$ and $V_{\rm rest}$. As mentioned above, by design of the K-focusing process, noise in the post-processed spectra can become correlated signal, for higher geometric correlations. Assuming that Poisson (white) noise is random for each spectrum, we can test this effect by splitting the data-set into independent samples. Correlated white noise will thus be different in each sample, allowing to estimate global effective uncertainties by measuring the CCF position of the CO signal for each sample. Different to the first test, this will measure the global uncertainties for all data points in the CCF map (independent from their geometric correlation).

We split the CRIRES observations in three samples by selecting random data from each night (without repetition). In this way, each data sample consists of different measurements, but covers similar orbital phases.
Then, we derive the CCF maps in the same way we did before for the whole data-set. As we show in Fig.~\ref{fig:TauBoo_samples}, the CO signal is detected for each sample at about $3\,{\rm \sigma}$ significance. We measure the position of the CO signal position using a least-squares fit with {\tt Saltire}. Similarly to the first test, we derive the RMS scatter of the position parameters. 

We find the resulting RMS scatter for $K_{\rm c}$ is $2.39\,{\rm km\,s^{-1}}$ and for $V_{\rm rest}$ $0.48\,{\rm km\,s^{-1}}$. These are about a factor of ten larger, compared to the fit uncertainties, we found earlier from the MCMC fit of the whole data-set. We also repeat the one dimensional Gaussian fit for each of the samples and find for $K_{\rm c}$, similar uncertainties for the returned RMS. In Tab.~\ref{tab:fit_TauBoo}, we add these systematic uncertainties for $V_{\rm rest}$ and $K_{\rm c}$ in quadrature to the MCMC uncertainties.

Summarising both of the above tests for systematic uncertainties, we first find clear systematics, when using data with different noise budget, but negligible systematics, when analysing the combined CCF map for correlation length scales which are smaller or similar than detectable the CO signal itself.

We conclude that, first, systematics indeed originate from correlated noise, produced by the K-focusing process. This noise can be caused by white noise, noise from data post-processing, but also from underfitted signal (See Sec.~\ref{wiggles}). Second, the correlation length scale in the $K_{\rm c}$ range is similar or larger than the CO signal itself, thus, must lead to underestimated uncertainties when fitting the CCF map alone. 


\subsection{Phase dependent modelling}

\begin{figure*}
	\includegraphics[width=\linewidth]{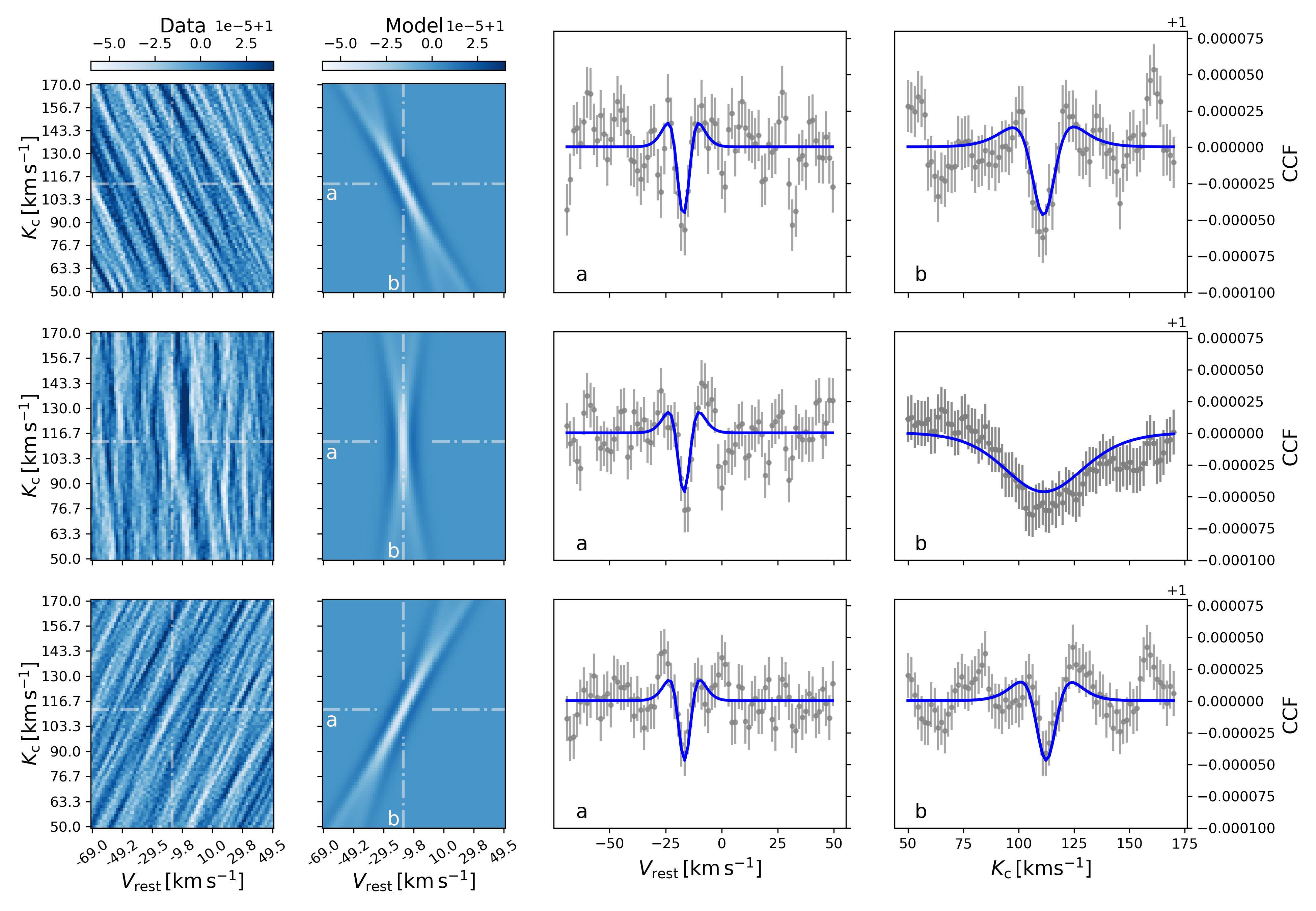}
    
    \caption{left panels, CO CCF-maps of $\rm \tau$ Bo\"otis b, reproduced from B12 data for different observation phases. from top to bottom: before, during, and after superior conjunction. Dashed lines indicate the position of maximum CCF contrast of the combined CCF map; Second panel from left: {\tt Saltire} models, for similar observation phases, evaluated from best fitting parameters of the combined fit; Second panel from right:
    Cuts trough the CCF-maps for the semi-amplitude of maximum CCF contrast of the combined CCF map. Gray dots: CCF-map data. Error-bars are estimated as the MCMC jitter term of the combined sample times $\rm \sqrt{3}$. Blue line: {\tt Saltire} model evaluated from best fitting parameters of the combined fit. Right panel: Similar cuts trough the CCF-maps, but for the rest velocity of maximum CCF contrast of the combined CCF map. }
    \label{fig:TauBoo_nights}
\end{figure*}

We use the best fit parameters in Tab.~\ref{tab:fit_TauBoo} to evaluate the {\tt Saltire} model separately for the phases of the three CRIRES observations at orbital phases before, during and after superior conjunction. Fig.~\ref{fig:TauBoo_nights} shows the CCF-maps for each observation, as well as the resulting models. The dependence of the CCF signal from the orbital phase is clearly visible. Despite not being a fit to the data, the model matches these observations rather well, with the absorption spectrum clearly detected in each observation. 

Assuming a stationary CO spectrum of the planet atmosphere, we would expect aliases originating from wiggles (here, spurious correlations of the CO line-mask with different parts of the CO spectrum, or with other atmospheric lines with similar patterns), which are quasi stationary at each observed phase (see Sec.~\ref{wiggles} for a discussion). This assumption can be used as a tool to disentangle wiggles from other aliases from correlated noise or imperfect post-processing of the data by simply comparing residual aliases of observations at different orbital phases and including quasi stationary components into the {\tt Saltire} model.

Fig.~\ref{fig:TauBoo_nights}, shows cuts (a) trough the CCF-maps for the semi-amplitude of maximum CCF contrast (of the combined CCF map). The detection significance (~3\,$\rm \sigma$) for each CRIRES observation does not allow to securely disentangle the origin of the aliases, which nevertheless,  will be possible as soon as higher signal-to-noise data will become available.

Measuring the Keplerian semi-amplitude from atmospheric signals obtained near phases of transits or eclipses introduces large uncertainties \citep[e.g.][]{Brogi18}. This can be understood by the middle panel of Fig.~\ref{fig:TauBoo_nights}, which shows the data during phases of superior conjunction. The CCF signal for both - data and model - show a very small correlation with the rest velocity, but cover a wide range of semi-amplitudes (see cut b). Measuring the mean position of the CCF signal by only evaluating the maximum detection significance can lead to large deviations produced by small correlated noise components along the chosen $V_{\rm rest}$ axis. Using {\tt Saltire} has clearly the advantage of being less biased to such small aliases, by taking the whole shape into account.

\section{Good practices}

Based on our modelling and the tests we described in Sec.~\ref{uncertainties}, we recommend following considerations when creating cross-correlation maps and fitting them with {\tt Saltire}:

\begin{itemize}
    
    \item Spectra need to be corrected from trends of the systemic velocity before creating the CCF map.
    
    \item For the $V_{\rm rest}$ range of the CCF map use the pixel resolution of the spectrograph as finest sampling to avoid additional correlations of the data. The sampling in the $K_{\rm c}$ range
    does not affect the results, as long as the CCF detection is well sampled.
    
    \item As for any least-square algorithm, choose your starting parameters and prior bounds close to the expected values. If you have clear side lobes, choose a value of $\Delta$ close to -0.5. Allowing -1 will increase the computational time significantly.
    
    \item If fitting for $h$, allow for a proper range. Too large a range will result in no improvement of the least-squares fit, with no errors being calculated. This will not affect MCMC results.
    
    \item The returned uncertainties are precision errors only. Systematic changes of the peak position due to correlated noise of the data are likely. Thus, they also need to be taken into account (e.g. by measuring independent data samples) when reporting fit uncertainties.
    
    \item A word of caution: As we described in Sec.~\ref{uncertainties}, noise is correlated by the K-focusing process. In other words, any noise structure in the CCF map can be misstaken as signal and fitted using the {\tt Saltire} model. Especially for very low detection significance ($< 3\,{\rm sigma}$), fitting CCF maps could result in a best fit for a correlated noise structure. Again, measuring independent data samples can help to verify such detections.
\end{itemize}

\section{Discussion}

HRCCS observations are widely used to detect and to map atomic and molecular species in exoplanet atmospheres both in the semi-amplitude and rest velocity range. A precise measurement of the companion's signal in $K_{\rm c} - V_{\rm rest}$ CCF-maps allows (i) to constrain the orbit of the companion, and thus, to turn a giant exoplanet or low-mass binary companion into a double-lined system, which allows to derive dynamical masses of both components, (ii) to constrain atmospheric dynamics by measuring the signal position of different species.
Measuring this position is difficult. It depends very much on the signal-to-noise, but also on the orbital phase observed. Transit transmission observations especially suffer from large uncertainties when measuring $K_{\rm c}$ solely from the position of the maximum CCF signal \citep{Brogi18}.

Aiming to improve the measurement of the CCF signals position we present {\tt Saltire}. An - easy to use - model to predict the phase dependent shape of CCF-maps, created for HRCCS observations of both exoplanet atmospheres, as well as of high-contrast binaries. The model allows to predict CCF-maps only by assuming a one dimensional line shape. We show that the two-dimensional shape for an CCF map results from a super positioning, for this one-dimensional function in the $K_{\rm c} - V_{\rm rest}$ frame, which we call K-focusing process. This is exactly the same process, which is in use for decades to generate CCF maps from HRCCS observations. Making use of the fact that the parameters that have been used to create the CCF map are fixed, we can use this model to efficiently fit complex two-dimensional shapes with a minimal set of parameters. In this work, we realise this one dimensional shape by a double Gaussian function. The {\tt Saltire} model is designed to facilitate any one-dimensional line shape (e.g. Lorentzian or more complex shapes). 

We demonstrate and evaluate the model for a simulated binary system, as well as for archival IR observations of the atmosphere of the gas giant exoplanet $\rm \tau$ Bo\"otis b. We use line-masks to create the CCF-maps, which allows us to derive average line profiles for the exoplanet. While the Gaussian approximation is expected to be a good fit to measure stellar spectra, we show that it can also be used to model the detection of atmospheric molecules in gas giants. By predicting the phase-dependent shape of an HRCCS observation, {\tt Saltire} can be used to plan HRCCS observations by simulating the combined CCF map expected from a certain observation strategy, but it can also be used to precisely fit the signal positing. 

With {\tt Saltire} we can recover the correct semi-amplitude with an accuracy better than $\rm 20\,m\,s^{-1}$, for a simulated noiseless observation. The companions's own signal ($\sim100\,\rm km\,s^{-1}$) is typically of order 5\,000 times greater than our method's accuracy, giving the potential to measure very accurate and precise dynamical masses. We also show that this accuracy can be achieved independently from the phase coverage of the observation, which allows to measure the semi-amplitude 10 times more accurate compared to a simple one-dimensional Gaussian fit, which is often used in literature. This phase independence of {\tt Saltire} is specially important for exoplanet observations, which usually cover only a part of the planet's orbit, resulting in asymmetric shapes of the CCF signal.

We show on simulated observations that in the absence of noise and post-processing aliases, the main source of uncertainties are aliases from spurious correlations with other lines/sets of lines within the spectrum (wiggles). With {\tt Saltire} we currently underfit the CCF signal, by not fitting for these aliases in the CCF. 

Including them in {\tt Saltire} would allow to better model the residual structure of high signal-to-noise CCF-maps and improve the parameter determination, especially for data with a short wavelength coverage, which typically enhances such correlations. In depth analysis of these wiggles, as well as possible methods to model them have recently been discussed for the detection of individual ionised, atomic species in the atmosphere of KELT-9b \citep{Borsato23}, as well as for high-precision RV measurements of double-lined binaries \citep{Lalitha23}. We show how such wiggles are represented in high signal to noise CCF maps due to the K-focusing process, and discuss the importance of including a phase dependent modelling of these structures when deriving high precision parameters. We propose that {\tt Saltire} can therefore serve as a diagnostic tool to disentangle wiggles from other aliases, e.g. due to post-processing of data for high signal to noise data.

For the detection of CO in the atmosphere of $\rm \tau$ Bo\"otis b (~5\,$\rm \sigma$ detection) we recover the semi-amplitude better than 1\,$\rm \sigma$, compared to the literature values. We achieve uncertainties of $2.4\,\rm km\,s^{-1}$ for the planet's semi-amplitude and $0.5\,\rm km\,s^{-1}$ for the rest velocity. 

We find that the main source of uncertainty is correlated noise, which is a by product of the K-focusing process and thus, is typically present in HRCCS CCF maps. This noise can be caused by white noise, data post-processing, or by wiggles. The estimated uncertainties $\rm \tau$ Bo\"otis b are on the order of 10 times larger than the fit errors. This is because, different to statistical noise in the CCF map, correlated noise becomes a CCF signal itself, thus fitting CCF maps will result in fit errors, which are underestimated. 

We show that it is possible to estimate the impact of correlated white noise by analysing independent, partial data-sets using {\tt Saltire}. That this is even possible for the CO detection in $\rm \tau$ Bo\"otis b, validates this approach to be extended to other low signal-to-noise detections from exoplanet atmospheres. This shows that {\tt Saltire} is an important tool to constrain the signal positions in CCF-maps from HRCCS observations. We further show that {\tt Saltire} is able to robustly return precise semi-amplitudes for different samplings in the $K_{\rm c}$ range of the CCF map offering possibilities to lower the computational costs when deriving CCF maps. 

We compare our results with {\tt Saltire} to a simple one-dimensional Gaussian fit of the same CCF map. Since both methods are limited by the impact of correlated noise, no substantial improvement can be achieved. Nevertheless, as we can clearly show for our noise-less simulations, {\tt Saltire} will allow to derive much more precise parameters for high signal-to-noise HRCCS detections.


Using observations covering a significant portion of the planet's orbit, phase resolved atmospheric properties can be retrieved from Ultra Hot Jupiters which offer a large signal to noise ratio. This has been successfully demonstrated for extremely irradiated planets like HD\,209458\,b \citep{Beltz21} and KELT-9b \citep{Pino22}. In {\tt Saltire}, we implement a weighting for the contribution of each observation. While this is used to account for uncertainties in the data, in the future {\tt Saltire} could be used to simulate the phase-dependent contribution of single observations to combined CCF-maps.


\section*{Acknowledgements}

This research is funded from the European Research Council (ERC) through the European Union’s Horizon 2020 research and innovation programme (grant agreement n$^\circ$803193/BEBOP). 
MB acknowledges partial support from the STFC research grant ST/T000406/1

\section*{Data Availability}

Most data underlying this article is available online as indicated in the specific section or reference. Data obtained with ESO telescopes are available in the ESO Science Archive Facility, at \url{http://archive.eso.org/cms.html}. Corrected data or CCF-maps, underlying this article will be shared on reasonable request to the corresponding author. Modelled data, can be generated using the {\tt Saltire} code, available on \href{https://github.com/dsagred/saltire}{Github}.



\bibliographystyle{mnras}
\bibliography{library} 




\appendix

\section{MCMC samples for simulated high-contrast binary and $\rm \tau$ Bo\"otis b observations.}

\begin{figure*}
	
	\includegraphics[width=\linewidth]{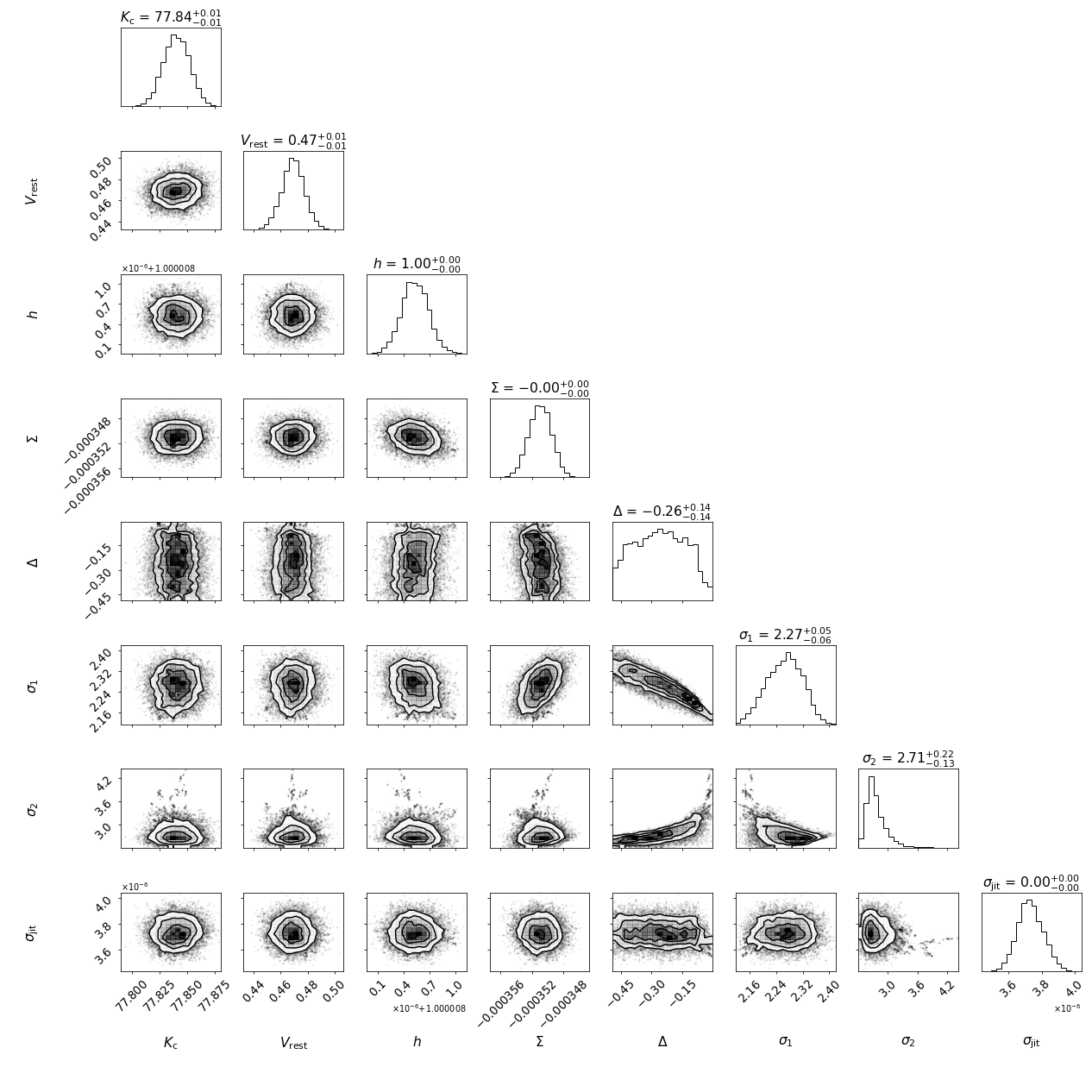}
    \caption{Corner plot of MCMC samples for {\tt Saltire} fit to CCF-map of simulated EBLM binary observation (Obs1).}
    \label{fig:Obs1_MCMC}
\end{figure*}

\begin{figure*}
	\includegraphics[width=\linewidth]{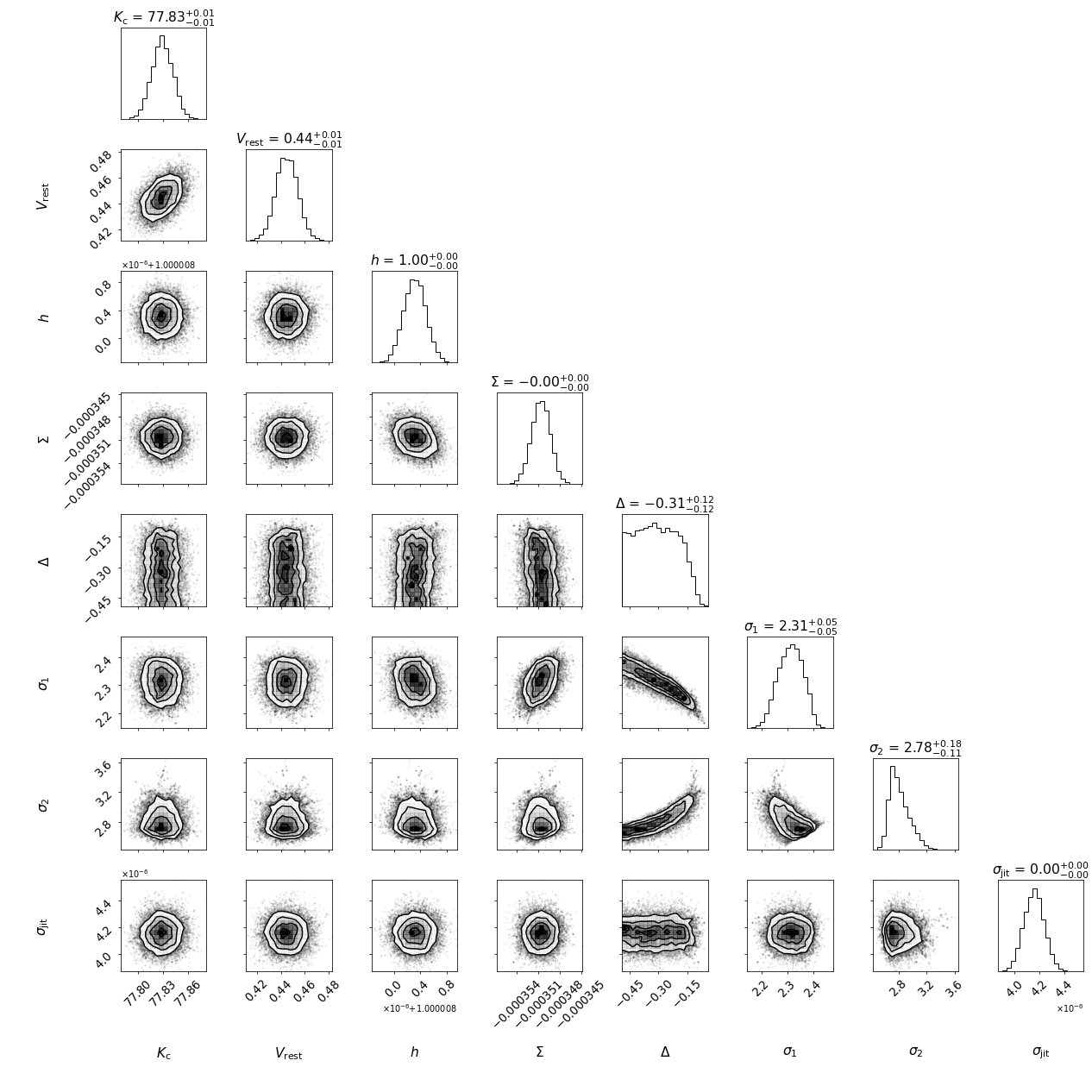}
    \caption{Corner plot of MCMC samples for {\tt Saltire} fit to CCF-map of simulated EBLM binary observation (Obs2).}
    \label{fig:Obs2_MCMC}
\end{figure*}

\begin{figure*}
	\includegraphics[width=\linewidth]{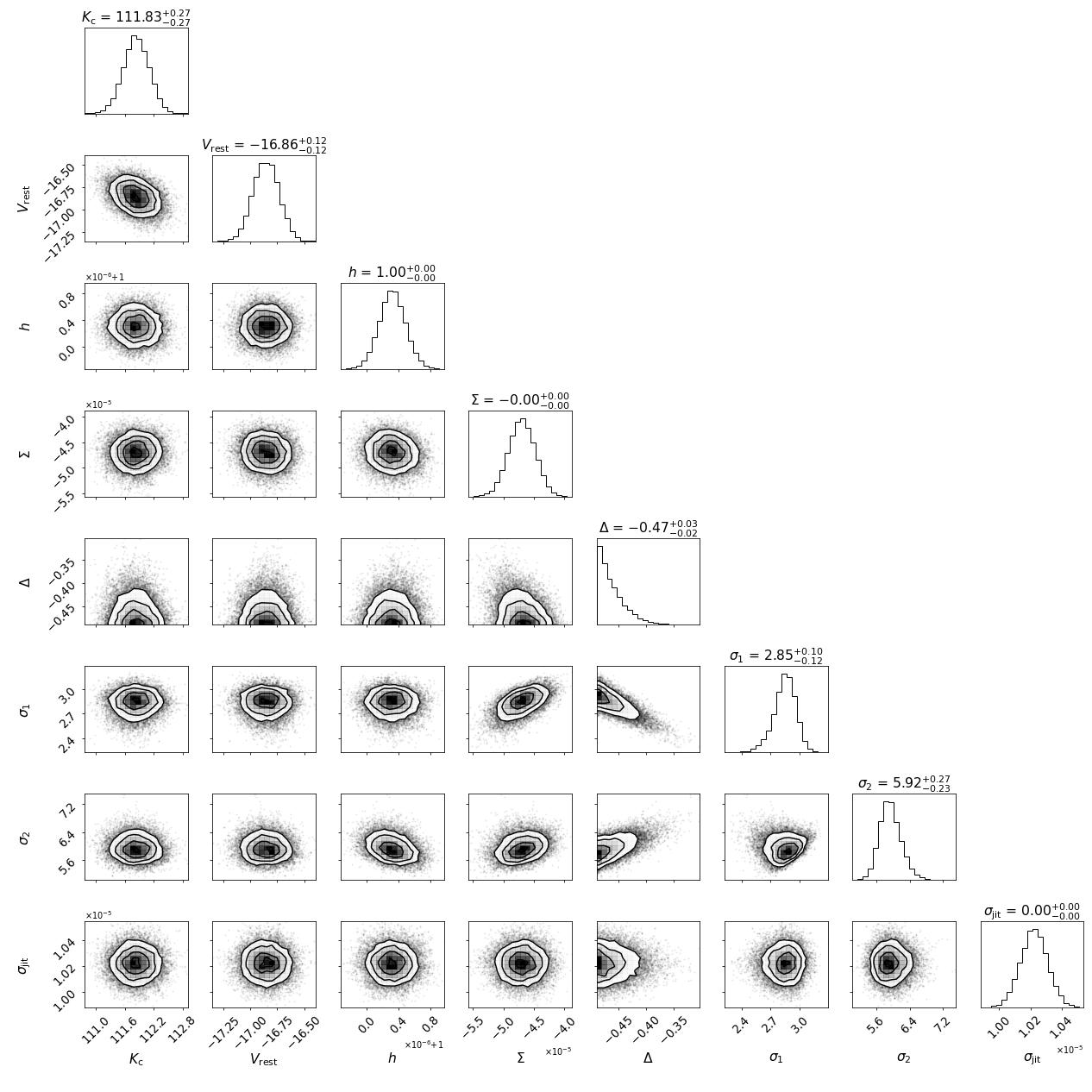}
    \caption{Corner plot of MCMC samples for {\tt Saltire} fit to combined CCF-map of $\rm \tau$ Bo\"otis b.}
    \label{fig:TauBoo_MCMC}
\end{figure*}

\section{Partial observations of $\rm \tau$ Bo\"otis b.}

\begin{figure*}
	\includegraphics[width=\linewidth]{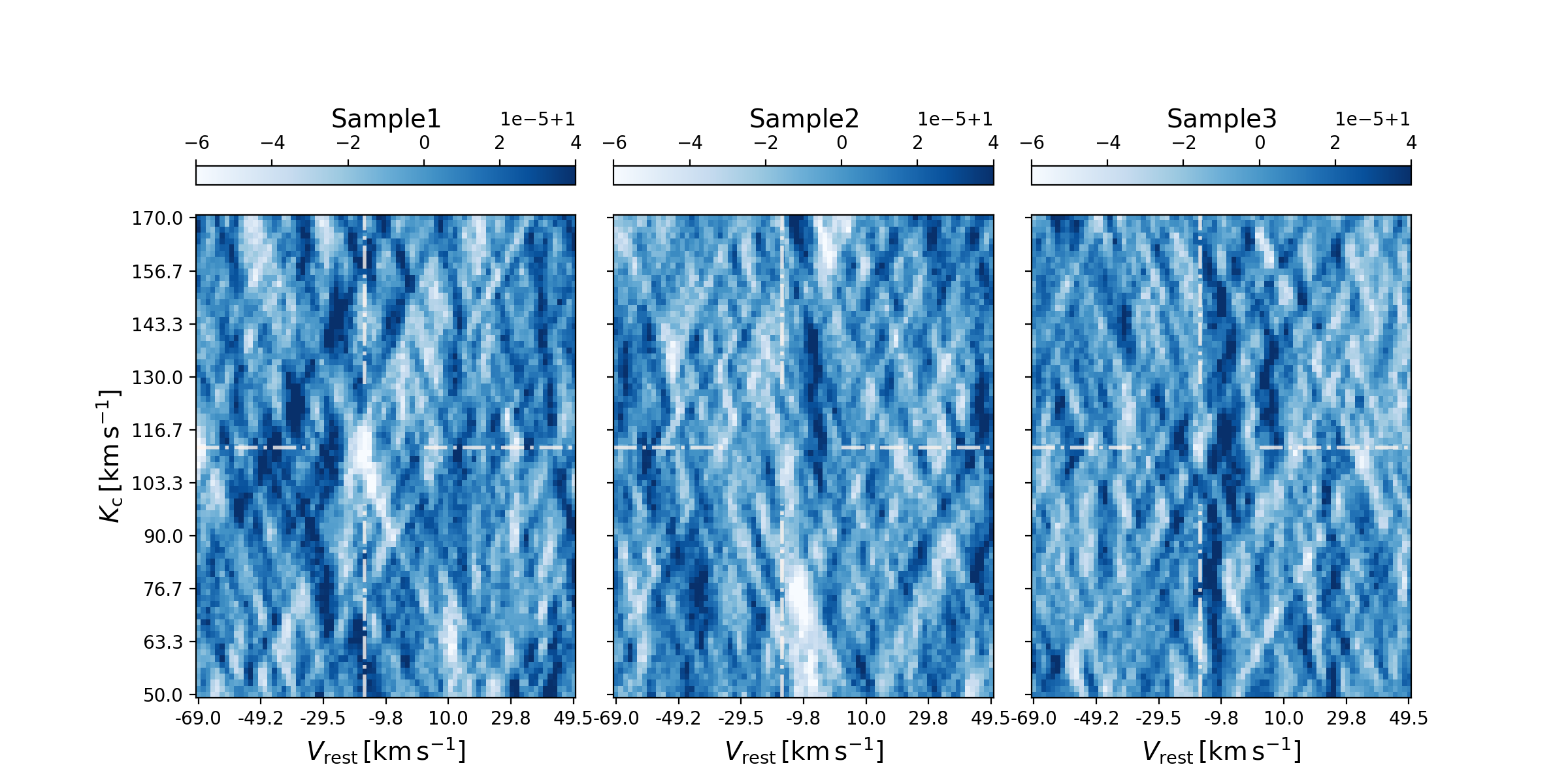}
    \caption{Combined CCF-maps for $\rm \tau$ Bo\"otis b. Each sample consists of different spectra, covering a similar orbital phase of the planet. White dashed lines mark the position of the CO signal. Correlated white noise creates unique structures for each of these CCF maps, sometimes even stronger than the CO signal itself.}
    \label{fig:TauBoo_samples}
\end{figure*}


\bsp	
\label{lastpage}

\end{document}